\documentclass[twocolumn,prx,aps,superscriptaddress]{revtex4-2}
\usepackage{amsmath}
\usepackage{amssymb}
\usepackage{amsfonts}
\usepackage{amssymb}
\usepackage{graphicx}
\usepackage{subfigure}
\usepackage{txfonts}
\usepackage{bm}
\usepackage{makeidx}
\usepackage{color}
\usepackage{mathtools}
\usepackage{threeparttable}
\usepackage{multirow}

\usepackage[colorlinks,linkcolor=blue,anchorcolor=blue,citecolor=blue,urlcolor=blue]{hyperref}

\begin{document}

\title{Quantum field theory for spin operator of the photon}

\author{Li-Ping Yang}
\affiliation{Center for Quantum Sciences and School of Physics, Northeast Normal University, Changchun 130024, China}
\affiliation{Birck Nanotechnology Center, School of Electrical and Computer Engineering, Purdue University, West Lafayette, IN 47906, U.S.A.}

\author{Farhad Khosravi}
\affiliation{Department of Electrical and Computer Engineering, University of Alberta, Edmonton, Alberta T6G 1H9, Canada}

\author{Zubin Jacob}
\affiliation{Birck Nanotechnology Center and Purdue Quantum Science and Engineering Institute, School of Electrical and Computer Engineering, Purdue University, West Lafayette, IN 47906, U.S.A.}
\email{zjacob@purdue.edu}

\begin{abstract}

All elementary particles in nature can be classified as fermions with half-integer spin and bosons with integer spin. Within quantum electrodynamics (QED), even though the spin of the Dirac particle is well defined, there exist open questions on the quantized description of spin of the gauge field particle---the photon. Using quantum field theory, we discover the  quantum operators for the spin angular momentum (SAM) $\boldsymbol{S}_{M}=(1/c)\int d^{3}x\boldsymbol{\pi}\times\boldsymbol{A}$ and orbital angular momentum (OAM) $\boldsymbol{L}_{M}=-(1/c)\int d^{3}x\pi^{\mu}\boldsymbol{x}\times\boldsymbol{\nabla}A_{\mu}$ of the photon, where $\pi^{\mu}$ is the conjugate canonical momentum of the gauge field $A^{\mu}$. We also reveal a perfect symmetry between the angular momentum commutation relations for Dirac fields and Maxwell fields. We derive the well-known OAM and SAM of classical electromagnetic fields from the above defined quantum operators. Our work shows that the \textcolor{red}{full canonical} spin and OAM operators commute\textcolor{red}{, whereas their transverse-field observable counterparts generally do not; nevertheless, the observable total angular momentum obeys the standard angular momentum algebra}. The correct commutation relations of orbital and spin angular momentum of the photon \textcolor{red}{have} applications in quantum optics, topological photonics as well as nanophotonics and can be extended in the future for the spin structure of nucleons.
\end{abstract}

\maketitle

\section{Introduction}
Spin is the fundamental property that distinguishes the two types of elementary particles: fermions with half-integer spin and bosons with integer spin. Beth's seminal experiment~\cite{beth1936mechnical} has shown that each circularly polarized plane-wave photon carries angular momentum of $\hbar$.  An earlier experiment work implemented by Raman and Bhagavantam even pointed out that this angular momentum belongs to the photon spin~\cite{raman1931experimental}. The polarizations of the electromagnetic (EM) field are commonly accepted as the spin degrees of the freedom of the photon. However, apart from these well established global properties of polarization,  more recently, the photon spin density, a local quantity which is a function of space and time has risen to the forefront of multiple fields~\cite{buttner2015dynamics,rodriguez2013near,petersen2014chiral,devlin2017arbitrary,gong2018nanoscale}.  It should be noted that a complete quantum treatment of the photon spin which connects space-time dependent fields and the global observables has never been achieved. 


This problem is of interest in quantum optics, nanophotonics and topological photonics. A substantial body of work based on the free-space classical Maxwell equations has been devoted to finding the measurable photon spin angular momentum (SAM) and orbital angular momentum (OAM)~\cite{van1994spin,van1994commutation,berry2009optical,barnett2010rotation,barnett2016natures,bialynicki2011canonical,Calvo2006quantum,Coles2012chirality,bliokh2015transverse}. Quantization of photon spin is also the signature of topological electromagnetic phases of matter~\cite{mechelen2018gyroelectric,soskin2016singular,barik2018topological} and skyrmion texture in optical scattering experiments~\cite{tsesses2018optical}. In the near-field of nanophotonic structures, evanescent waves exhibit universal spin-momentum locking widely studied in 2D materials, photonic crystal waveguides, optical fibers and metamaterials ~\cite{bliokh2014extraordinary,van2016universal,lodahl2017chiral,aiello2015transverse,stav2018quantum}. Here, an advancement for these fields is reported by exploiting a paradigm shift in approach for photonics---we appeal to a fundamental QED lagrangian including Dirac particles to quantize the spin of the light field.

Even in the context of high-energy physics, there is an on-going discussion on the decomposition of the angular momentum of the photon or gluon into SAM and OAM parts ~\cite{belinfante1939spin,jaffe1990g1,Ji1997gauge,chen2008spin,Wakamatsu2010gauge,Lorce2013Geometrical,gong2018nanoscale}. Leader and Lorc\'{e} have written a pedagogical review to explain these important open challenges for the field~\cite{leader2014angular}. The fundamental difficulty stems from the puzzling fact that the genuine gauge-invariant photon spin operator does not exist. Our work utilizes quantum field theory to pave the way and resolve these questions about photon spin with future implications for the spin structure of the nucleon~\cite{ashman1989investigation}. 

\begin{figure*}[hbt!]
\centering
\includegraphics[width=17cm]{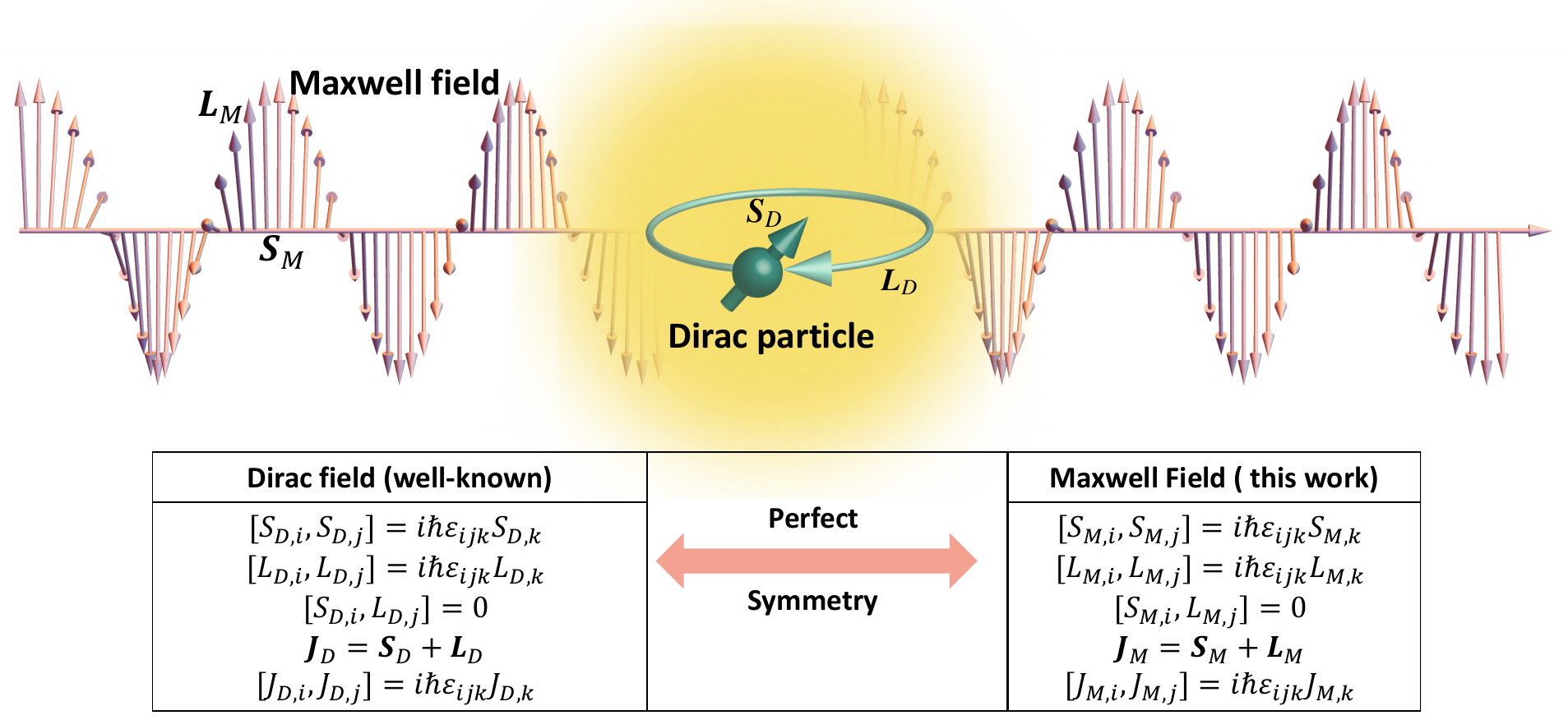}
\caption{\label{fig1} \textbf{Comparison between our proposed photon angular momentum operators and the well-known Dirac field counterparts.} We show that our discovered quantum operators $\boldsymbol{S}_M$ and $\boldsymbol{L}_M$ obey the canonical commutation relations in striking parallel to Dirac fermions.}
\end{figure*}

The importance of the problem becomes clear on comparing to the  Dirac spin operator $\boldsymbol{S}_{D}=(\hbar/2)\int d^{3}x\psi^{\dagger}\hat{\boldsymbol{\Sigma}}\psi$. These obey the canonical commutation relationships for angular momenta.  However, a genuine quantum operator for the photon spin $\boldsymbol{S}_M$, which satisfy the standard equal-time commutation relations $[S_{M,i},S_{M,j}]=i\hbar\varepsilon_{ijk}S_{M,k}$, have never been obtained. Because of this major knowledge gap, interesting questions have been raised whether photon spin and OAM are true observable angular momenta at all~\cite{van1994commutation,van1994spin,barnett2016natures,Arnaut2000orbital}. We solve this open problem within the canonical quantization framework of relativistic field theory. We explicitly derive the quantum operators for the spin $\boldsymbol{S}_M$ and OAM $\boldsymbol{L}_M$ of the photon by quantizing the electromagnetic (EM) field covariantly in the Lorenz gauge. Using relativistic field theory, we show a perfect symmetry between the angular momentum commutation relations for Dirac fermions and Maxwell bosons. We show that photon SAM and OAM operators commute with each other and are thus separately observable.  This is an advance over existing knowledge since it is an important question whether the spin and OAM of light are independently observable ~\cite{van1994commutation,van1994spin}.

Historically, duality symmetry of classical electromagnetism (Maxwell's equations) in source free regions was the chosen route to understand conservation relation between the photon  spin density and helicity ~\cite{candlin1965analysis,calkin1965invariance,lipkin1964existence}. Important recent works have generalized this approach ~\cite{cameron2012electric,drummond1999dual}. However, QED requires local U(1) gauge symmetry, to capture quantum light-matter interaction.  Our work also fills this historic gap by investigating the angular-momentum conservation law of the  combined Dirac-Maxwell fields. An important hallmark of our QED approach is that we obtain the well-known Dirac SAM and OAM operators simultaneously en route to our new photonic spin and OAM operators. We note that similar to the Aharonov-Bohm phase~\cite{Marletto2020ABphase} where gauge fields and not electric/magnetic fields take center stage, the quantum spin operator of light also shares this characteristic of gauge field theory.    

In this paper, we prove the perfect symmetry between the angular momentum of Dirac fermions and Maxwell bosons using quantum field theory. For the broad audience, we summarize our central result in a simple schematic (Figure~\ref{fig1}). First, we report on the discovery of new SAM and OAM quantum operators for the photon. While the commutation relations of angular momenta for Dirac operators are well established, we show here that Maxwell operators obey the same striking symmetry. Our discovered operators for the photon spin ($\boldsymbol{S}_M$) necessarily requires the inclusion of virtual photons in QED. Table~\ref{tab:dual} shows a summary of our theoretical formalism that includes SO(3) rotational symmetry and local U(1) gauge symmetry of QED. On the other hand, the important previous framework of duality symmetry in Maxwell's equations \cite{barnett2016natures} only deals with the local spin density of transverse photons. Duality symmetry in Maxwell's equations does not capture the global photon spin which includes real (transverse) and virtual (longitudinally polarized in vacuum) photons.  In this paper, we only focus on the spin of the photon. However, we believe our results can be generalized in the future to the other massless gauge boson---the gluon~\cite{leader2014angular,lowdon2014boundary}.

The perfect symmetry between Dirac and Maxwell angular momentum operators necessarily includes the subtle role of longitudinal and scalar photons which are not gauge invariant. To incorporate gauge invariance into the theoretical framework, we put forth a  re-decomposition of the total angular momentum of the combined Dirac-Maxwell fields (see Tab.~\ref{tab:gauge}). We reveal that the contribution of the photon spin from longitudinal photons is hidden from detection by the requirement of gauge invariance. The experimentally measurable part of the photon spin is its transverse-field part. The gauge field necessarily includes longitudinal and transverse fields as required by relativistic field theory - QED. {\color{red}This leads to nonstandard commutation relations for the separate observable photon SAM and OAM, while their sum satisfies the standard angular momentum algebra for free radiation $[J^{\rm obs}_{M,i},J^{\rm obs}_{M,j}]=i\hbar\varepsilon_{ijk}J^{\rm obs}_{M,k}$.} In our manuscript, we use the super-script ``OBS" as opposed to the notation of gauge invariant variables. We note that the SAM and OAM operators commute. We prove that this is true for the new gauge operators as well as the well-known OAM and SAM operators. These results have implications for future experiments on photon spin noise~\cite{Ballantinee2016spinnoise} and exotic topological phases of matter~\cite{todd2019nonlocal}.

\begin{table*}
\centering
\setlength{\tabcolsep}{4pt}
\footnotesize
\begin{tabular}{|p{60pt}||p{130pt}||p{125pt}|p{130pt}|}
\hline
 & Current approach~\cite{candlin1965analysis,calkin1965invariance,lipkin1964existence,cameron2012electric,drummond1999dual}
 & \multicolumn{2}{c|}{Our work}\\
\hline
Symmetry type
& Duality symmetry
& $\mathrm{SO}(3)$ rotational symmetry
& Local $\mathrm{U}(1)$ gauge symmetry\\
\hline
Symmetry transformation
& $\boldsymbol{E}\rightarrow\boldsymbol{E}\cos\theta
+c\boldsymbol{B}\sin\theta$
\par
$\boldsymbol{B}\rightarrow\boldsymbol{B}\cos\theta
-(\boldsymbol{E}/c)\sin\theta$
& $\boldsymbol{x}\rightarrow
R(\boldsymbol{\theta})\boldsymbol{x}$
\par
($R$: $3\times 3$ matrix)
\par
$\psi(\boldsymbol{x})\rightarrow
e^{i\boldsymbol{J}\cdot\boldsymbol{\theta}}
\psi(\boldsymbol{x})$
& $A_{\mu}(x)\rightarrow
A_{\mu}(x)-\partial_{\mu}f(x)$
\par
$\psi(x)\rightarrow
\psi(x)e^{iqf(x)/\hbar}$\\
\hline
Physical phenomenon
& Conservation of helicity
& Conservation of angular momentum
& Massless photon\\
\hline
Involved spin angular momentum
& Provides local classical spin density of the transverse EM field; does not lead to angular momentum commutation relations
& Full spin operator of the isolated photon [U(1) gauge field] satisfies the correct commutation relations
& Introducing interaction between Dirac--Maxwell fields leads to the gauge-invariant photon spin\\
\hline
\end{tabular}
\caption{\label{tab:dual}
\textbf{Our work is fundamentally beyond duality symmetry and incorporates SO(3) rotational symmetry and local U(1) gauge symmetries.}}
\end{table*}

\section{Historical context of the quantum spin operator of the photon}
To show the novelty of this work, we give a short review of the theoretical development of the photon spin in quantum optics. In Chap. I of the text book~\cite{cohen1997photons}, the authors have shown that the total angular momentum of classical electrodynamics (CED) can be split into three parts:
\begin{equation}
\boldsymbol{J}_{\rm CED}=\sum_{\alpha}\boldsymbol{x}_{\alpha}\times\boldsymbol{p}_{\alpha}+\boldsymbol{S}^{\rm obs}_M+\boldsymbol{L}^{\rm obs}_M. \label{eq:J_CED}
\end{equation}
Here, the first term denotes the OAM of charge particles and $\boldsymbol{p}_{\alpha}=m_{\alpha}\dot{\boldsymbol{x}}_{\alpha}+q_{\alpha}\boldsymbol{A}_{\perp}$ is the canonical momentum of the $\alpha$th particle with charge $q_{\alpha}$ and mass $m_{\alpha}$. The subscript $\ _{\perp}$ of the vector potential denotes its transverse part with vanishing divergence, i.e., $\boldsymbol{\nabla}\cdot\boldsymbol{A}_{\perp}$=0. The second and third terms have been interpreted as the SAM and OAM of light, respectively,
\begin{align}
\boldsymbol{S}^{\rm obs}_{M} & = \varepsilon_0 \int d^3x \boldsymbol{E}_{\perp}\times\boldsymbol{A}_{\perp},\\
\boldsymbol{L}^{\rm obs}_{M} & = \varepsilon_0 \int d^3x E_{\perp}^j(\boldsymbol{x}\times\boldsymbol{\nabla})A_{\perp}^j.
\end{align}
Here, we have added the superscript ``obs" in $\boldsymbol{S}^{\rm obs}_{M}$ and $\boldsymbol{L}^{\rm obs}_{M}$, because we will show in later sections these are only the directly observable part of the photonic angular momenta. The total angular momenta $\boldsymbol{J}_{\rm CED}$ of CED is a conserved quantity, and all the three parts in Eq.~(\ref{eq:J_CED}) are invariant under a classical gauge transformation. Thus, the decomposition of the angular momentum of CED in free space has been completely solved.

Two fundamental problems in electrodynamics angular momentum decomposition occur after quantization. Firstly, the OAM of a charged particle $\boldsymbol{p}=-i\hbar\boldsymbol{\nabla}$ is not gauge-invariant anymore because an extra space-time-dependent phase is acquired under a gauge transformation. Secondly, the early important work by van Enk and Nienhuis has shown that the photon spin from CED does not satisfy the angular momentum commutation relation~\cite{van1994commutation,van1994spin}, $[S^{\rm obs}_{M,i},S^{\rm obs}_{M,j}]=0$. Thus the open question remains whether new quantum operators exist beyond these well known classical results that satisfy the correct canonical commutation relations. In this paper, we 
find these SAM and OAM operators of photons using quantum field theory. Furthermore, we also derive the above well known classical decomposition through the quantum operators.

Due to these two unsolved problems and the lack of a quantum gauge theory, the conservation of photon spin has been studied ~\cite{barnett2016natures} only in the absence of charges  using electric-magnetic duality symmetry ~\cite{barnett2010rotation,cameron2012electric,bliokh2013dual}. By building on Lipkin-Calkin conservation law~\cite{lipkin1964existence,calkin1965invariance}, the photon spin density has been previously interpreted $\boldsymbol{s}^{\rm obs}(\boldsymbol{x},t)$ as the current corresponding to the photonic helicity $h(\boldsymbol{x},t)$, i.e., $\partial h/\partial t+\boldsymbol{\nabla}\cdot \boldsymbol{s}^{\rm obs}=0$. An experiment has also been implemented to measure the photonic helicity and the related quantity photonic chirality~\cite{forbes2021measures}. We note that these elegant works are of broad interest, but their results do not lead to new SAM and OAM operators especially for the gauge field. Electric-magnetic duality of Maxwell's equations is a classical symmetry which does not apply to U(1) gauge fields which are essential to QED. The duality symmetry only gives the conservation of the photon helicity, not the angular momentum of gauge fields. On the other hand, in the presence of charges (Dirac fermions), this duality symmetry will be destroyed. In QED, it has been well-accepted that the conservation of angular momentum is due to the SO(3) rotational symmetry of the background space-time. The duality-symmetry-based argument cannot be extended to explain the conservation of spin for all other particles (e.g., gluons and Dirac fermions). Therefore, we appeal to quantum field theory to discover new SAM and OAM operators and also present a unified framework for gauge-field SAM and OAM.

In the high-energy community, the separation of the total angular momentum of photons and gluons into gauge-invariant spin and orbital contributions is an important and interesting challenge faced by gauge field theories like QED and quantum chromodynamics (QCD)~\cite{leader2014angular}. Deriving the angular momentum commutation relations from the basic postulated relation between the field and its canonical momentum is also of fundamental significance. Here, we solve these two fundamental problems in QED angular momentum decomposition conclusively. Our work also leads to the derivation of the angular momentum in classical electrodynamics theory from the new operators discovered within quantum gauge-field theory.

\section{New OAM and SAM operators for the photon}

We utilize a quantum field theory framework to analyze the spin and orbital angular momentum of the photon. In quantum optics (non-relativistic QED), only the transverse degrees of freedom for the photon are quantized. In stark contrast, our relativistic treatment shows that longitudinally polarized photons are necessary to construct the full spin-1 operator for the photon. The subtle detail, overlooked previously, becomes self-evident in our starting Lagrangian that incorporates both the longitudinal part of the vector potential $\boldsymbol{A}$ and the scalar potential $A_0$. These quantities can not be quantized with the standard Maxwell Lagrangian density~$\mathcal{L}_{M,{\rm ST}}=-F^{\mu\nu}F_{\mu\nu}/4\mu_0$ ($F_{\mu\nu}=\partial_{\mu}A_{\nu}-\partial_{\nu}A_{\mu}$ is the  EM field tensor), because there is no canonical conjugate momentum corresponding to the scalar field $A^{0}$ and the longitudinal potential $\boldsymbol{A}_{\parallel}$ with zero curl ($\boldsymbol{\nabla}\times\boldsymbol{A}_{\parallel}=0$) has also been shown to be a redundant dynamical variable (see Chap. II in Ref.~\cite{cohen1997photons}
). 

To obtain complete knowledge of the polarization degrees of freedom, we start from the gauge-fixed Maxwell Lagrangian density~\cite{cohen1997photons,greiner2013field} 
\begin{equation}
\mathcal{L}_{M}=-(\partial_{\mu}A^{\nu})(\partial^{\mu}A_{\nu})/2\mu_{0}. \label{eq:LagrangianM} 
\end{equation}
The covariant quantization of the photon in the Lorenz gauge can be realized by defining the canonically conjugate momentum~\cite{cohen1997photons,greiner2013field} 
\begin{equation}
\pi_{\mu}=\frac{\partial\mathcal{L}_{M}}{\partial(\partial_{0}A^{\mu})}=-\frac{1}{\mu_{0}}\partial^{0}A_{\mu},
\end{equation}
and postulating the fundamental equal-time commutation relations (ETCRs),
\begin{align}
[A^{\mu}(\boldsymbol{x},t),\pi^{\nu}(\boldsymbol{x}',t)] & =i\hbar cg^{\mu\nu}\delta^{3}(\boldsymbol{x}-\boldsymbol{x}'),\label{eq:ECTR1}\\
[A^{\mu}(\boldsymbol{x},t),A^{\nu}(\boldsymbol{x}',t)] & =[\pi^{\mu}(\boldsymbol{x},t),\pi^{\nu}(\boldsymbol{x}',t)]=0,\label{eq:ECTR2}
\end{align}
with the metric tensor $g^{\mu\nu}=\rm{diag}\{1,-1,-1,-1\}$ of the Minkowski space and the speed of light $c=1/\sqrt{\mu_0\varepsilon_0}$ in vacuum. The photon Hamiltonian is given by
\begin{equation}
H_M = -\frac{1}{2\mu_{0}}\int d^3x\left[\mu_0^2\pi^{\mu}\pi_{\mu}+(\boldsymbol{\nabla}A^{\mu})\cdot(\boldsymbol{\nabla}A_{\mu})\right].\label{eq:H_M}
\end{equation}
We note that $\pi_{\mu}$ and $A_{\mu}$ are now quantum operators. But, to highlight the spin degrees, we only add the $\hat{\ }$ symbol on the spin matrices throughout this paper. 

The fundamental connection between a continuous symmetry and and the corresponding conservation law was given by Noether. Applying Noether's theorem on the Lorentz rotation symmetry~\cite{greiner2013field}, we obtain the angular momentum tensor density from $\mathcal{L}_M$~\cite{van2000theory}(also see Appendix~\ref{sec:appendixA}),
\begin{align}
M_{M}^{\mu\nu\lambda} & =\Theta_{M}^{\mu\lambda}x^{\nu}-\Theta_{M}^{\mu\nu}x^{\lambda}+\frac{\partial\mathcal{L}_{M}}{\partial(\partial_{\mu}A^{\sigma})}(I^{\nu\lambda})^{\sigma\tau}A_{\tau}\label{eq:AMtensor1}\\
 & =\Theta_M^{\mu\lambda}x^{\nu}-\Theta_M^{\mu\nu}x^{\lambda}-\frac{1}{\mu_{0}}[(\partial^{\mu}A^{\nu})A^{\lambda}-(\partial^{\mu}A^{\lambda})A^{\nu}]\label{eq:AMtensor2}
\end{align}
where $\Theta_M^{\mu\lambda}$ is the energy-momentum tensor and the infinitesimal Lorentz transformation generator for the vector field is given by
\begin{equation}
(I^{\alpha\beta})^{\mu\nu}=g^{\alpha\mu}g^{\beta\nu}-g^{\alpha\nu}g^{\beta\mu},
\end{equation}
which is a an anti-symmetric matrix $(I^{\alpha\beta})^{\mu\nu}=-(I^{\beta\alpha})^{\mu\nu}$. The first two terms in Eq.~(\ref{eq:AMtensor2}) comes from the spatial coordinate rotation and the last term denotes the contribution from the ``intrinsic" rotation between different components of the vector potential $A_{\mu}$~\cite{greiner2013field}.

Focusing on the three-dimensional rotation symmetry, we, arrive at the central result of our paper---the striking quantum operators of the spin and OAM of the photon,
\begin{equation}
\boldsymbol{S}_{M}=-\frac{1}{\mu_{0}c}\int d^{3}x(\partial_{0}\boldsymbol{A})\times\boldsymbol{A}=\frac{1}{c}\int d^{3}x\boldsymbol{\pi}\times\boldsymbol{A} \label{eq:SM-lorenz}   
\end{equation}
and 
\begin{equation}
\boldsymbol{L}_{M}=\frac{1}{\mu_{0}c}\!\!\int\!\! d^{3}x(\partial_{0}A^{\mu})\boldsymbol{x}\times\boldsymbol{\nabla}A_{\mu}=-\frac{1}{c}\!\int\!\! d^{3}x\pi^{\mu}\boldsymbol{x}\times\boldsymbol{\nabla}A_{\mu}, 
\end{equation}
from the rotations of the intrinsic and spatial degrees of freedom, respectively. Of course, given the long-standing nature of the problem, fundamental checks are required to verify these are indeed the SAM and OAM of photons. Utilizing the ETCRs in Eqs.~(\ref{eq:ECTR1}) and (\ref{eq:ECTR2}), we show that our defined photon spin and OAM operators satisfy the standard angular momentum commutation relations (see Appendix~\ref{sec:appendixB} and also Appendix~\ref{sec:CR-recheck})
\begin{align}
[S_{M,i},S_{M,j}] & =i\hbar\varepsilon_{ijk}S_{M,k},\label{MCR1} \\
[L_{M,i},L_{M,j}] & =i\hbar\varepsilon_{ijk}L_{M,k},\label{MCR2}\\
[L_{M,i},S_{M,j}] & =0,\label{MCR3}
\end{align}
where $\varepsilon_{ijk}$ is the three-dimensional Levi-Civita tensor and $i,j = 1,2,3$. Deriving the commutation relations for the photon spin and OAM operators from the basic field ETCR ansatz (\ref{MCR1}) and (\ref{MCR2}) is of fundamental importance. This has never been achieved till date. Note the other striking result---the SAM and OAM operators of the photon commute. We also emphasize that this commutation relation can not be obtained from the standard Maxwell Lagrangian density $\mathcal{L}_{M,{\rm ST}}$ under the non-covariant quantization scheme. 

It is well known that the Dirac spin operators obey SU(2) symmetry. To clearly show the SO(3) symmetry in the quantum spin degrees of the photon, we perform the plane-wave expansions on the vector potential and its canonically conjugate momentum (see Chap. 7 in Ref~\cite{greiner2013field})
\begin{align}
A^{\mu}&=\!\!\int\!\! d^{3}k\!\sum_{\lambda=0}^{3}\!\!\sqrt{\frac{\hbar}{2\varepsilon_{0}\omega_{\boldsymbol{k}}(2\pi)^{3}}}\left[a_{\boldsymbol{k},\lambda}\epsilon^{\mu}(\boldsymbol{k},\lambda)e^{i\boldsymbol{k}\cdot\boldsymbol{x}}+\rm{h.c.}\right], \label{eq:Amu}\\
\pi^{\mu}&=i\!\!\int\!\! d^{3}k\!\sum_{\lambda=0}^{3}\!\!\sqrt{\frac{\hbar\omega_{\boldsymbol{k}}}{2\mu_{0}(2\pi)^{3}}}\left[a_{\boldsymbol{k},\lambda}\epsilon^{\mu}(\boldsymbol{k},\lambda)e^{i\boldsymbol{k}\cdot\boldsymbol{x}}-\rm{h.c.}\right],\label{eq:pimu}
\end{align}
where $\omega_{\boldsymbol{k}}=c|\boldsymbol{k}|$ is frequency of the mode with wave vector $\boldsymbol{k}$ and the unit vectors $\epsilon(\boldsymbol{k},\lambda)$ describe the four polarization photons. Following the convention~\cite{greiner2013field,cohen1997photons}, we let the two unit vectors $\epsilon(\boldsymbol{k},1)$ and $\epsilon(\boldsymbol{k},2)$ denote the two transverse modes, $\epsilon(\boldsymbol{k},3)=(0,\boldsymbol{k}/|\boldsymbol{k}|)$ for the longitudinal photon, and $\epsilon(\boldsymbol{k},0)=(1,0,0,0)$ for the scalar photon. In the following, we also use  $\boldsymbol{\epsilon}(\boldsymbol{k},\lambda)$ to denote the spatial part of the four-vector $\epsilon(\boldsymbol{k},\lambda)$. From the ETCR ansatz in Eq.~(\ref{eq:ECTR1}) and (\ref{eq:ECTR2}), we can derive the familiar bosonic commutation relations for the ladder operators $
[a_{\boldsymbol{k},\lambda},a_{\boldsymbol{k}',\lambda'}^{\dagger}]  =-g_{\lambda\lambda'}\delta^{3}(\boldsymbol{k}-\boldsymbol{k}')$ and $  [a_{\boldsymbol{k},\lambda},a_{\boldsymbol{k}',\lambda'}]  = [a_{\boldsymbol{k},\lambda}^{\dagger},a_{\boldsymbol{k}',\lambda'}^{\dagger}]=0$.

Using the plane-wave expansion, we now re-express our discovered photon spin operator (\ref{eq:SM-lorenz}) in an intuitive form in wave-vector space  (see Appendix~\ref{sec:plane-wave expansion})
\begin{equation}
\boldsymbol{S}_{M}=\hbar\int d^{3}k \phi^{\dagger}_{\boldsymbol{k}}\boldsymbol{\hat{s}}\phi_{\boldsymbol{k}}, \label{eq:SM_matrix}
\end{equation}
where the column-vector  $\phi_{\boldsymbol{k}}=[a_{\boldsymbol{k},1},a_{\boldsymbol{k},2},a_{\boldsymbol{k},3}]^{T}$ is the field operator of the photon in wave-vector space and the $3\times 3$ matrix $\boldsymbol{\hat{s}}=\sum_{\lambda=1}^3\hat{s}_{\lambda}\boldsymbol{\epsilon}(\boldsymbol{k},\lambda)$ is the spin-$1$ operator of the photon with the SO($3$) rotation generators
\begin{equation}
\hat{s}_{1}=\left[\begin{array}{ccc}
0 & 0 & 0\\
0 & 0 & -i\\
0 & i & 0
\end{array}\right],\ \hat{s}_{2}=\left[\begin{array}{ccc}
0 & 0 & i\\
0 & 0 & 0\\
-i & 0 & 0
\end{array}\right],\ \hat{s}_{3}=\left[\begin{array}{ccc}
0 & -i & 0\\
i & 0 & 0\\
0 & 0 & 0
\end{array}\right].
\end{equation}
Here, we see that our defined photon spin operator generates the rotation of the polarization degrees of freedom of light.

The direction of our defined photon spin is completely determined by the polarization [i.e., the unit vector $\boldsymbol{\epsilon}(\boldsymbol{k},\lambda)$] of the photon. Thus, the spin operator indeed describes the angular momentum carried by the polarization degrees of freedom of the photon. This is significantly different from the OAM of the photon that we obtain {\color{red}
\begin{equation}
\boldsymbol{L}_{M}
=i\hbar\int d^{3}k
\sum_{\lambda,\lambda'=0}^{3}
a_{\boldsymbol{k},\lambda}^{\dagger}
\left[
g^{\lambda\lambda'}
(\boldsymbol{k}\times\boldsymbol{\nabla}_{\boldsymbol{k}})
+\boldsymbol{\Gamma}_{\lambda\lambda'}(\boldsymbol{k})
\right]
a_{\boldsymbol{k},\lambda'},
\end{equation}
where
$\boldsymbol{\Gamma}_{\lambda\lambda'}(\boldsymbol{k})
=\epsilon^{\mu}(\boldsymbol{k},\lambda)
(\boldsymbol{k}\times\boldsymbol{\nabla}_{\boldsymbol{k}})
\epsilon_{\mu}(\boldsymbol{k},\lambda')$
accounts for the momentum dependence of the polarization frame.} In Appendix~\ref{sec:plane-wave expansion}, we prove that counter-rotating wave terms $a_{\boldsymbol{k},\lambda}a_{\boldsymbol{-k},\lambda'}$ and $a^{\dagger}_{\boldsymbol{k},\lambda}a^{\dagger}_{\boldsymbol{-k},\lambda'}$ both in the spin and OAM operators vanish since they change their sign when we relabel the indices $\{\boldsymbol{k},\lambda\}\rightarrow\{-\boldsymbol{k},\lambda'\}$. We also note that different from the spin of Dirac field, the photon spin is not the intrinsic angular momentum of light, because the polarization unit vectors $\boldsymbol{\epsilon}$ are dependent on the wave vector $\boldsymbol{k}$.

There remain two subtle aspects that need further exploration for developing a full quantum theory of photon spin. Firstly, there is a fundamental requirement in QED that a measurable quantity can not change under a gauge transformation. However, both $\boldsymbol{S}_{M}$ and $\boldsymbol{L}_{M}$ defined above for the free-space photon are not gauge invariant because longitudinal and scalar photons are involved. Thus, they are not direct physical observables. We argue that this is a fundamental tenet in the construction of the correct quantum theory because additional hidden degrees of freedom are necessary to construct the above quantum spin-1 operator for the free-space photon. On the other hand, only two transverse polarizations are allowed for the photon in free space. Secondly, in the presence of charges (Dirac particles), the EM field acquires a longitudinal (near-field) component that is beyond the transverse photons commonly encountered in vacuum. Can we construct a gauge-invariant photon spin operator in the presence of charges? Next, we will answer this question conclusively and show how to incorporate the gauge invariance into the photon angular momenta.

\begin{table*}[t]
\centering
\footnotesize
\setlength{\tabcolsep}{4pt}
\renewcommand{\arraystretch}{1.45}
\begin{tabular}{|p{60pt}|p{75pt}|p{105pt}|p{85pt}|p{105pt}|}
\hline
 & {\centering Dirac SAM\par}
 & {\centering Dirac OAM\par}
 & {\centering Maxwell SAM\par}
 & {\centering Maxwell OAM\par} \\
\hline
{\centering Canonical decomposition\par}
 & $\begin{gathered}
 \boldsymbol{S}_D=\int d^3x \psi^{\dagger}\hat{\boldsymbol{\Sigma}}\psi
 \end{gathered}$
 & $\begin{gathered}
 \boldsymbol{L}_D=\int d^3x \psi^{\dagger}(\boldsymbol{x}\times\boldsymbol{p})\psi
 \end{gathered}$
 & $\begin{gathered}
 \boldsymbol{S}_M=\frac{1}{c}\int d^3x \boldsymbol{\pi}\times\boldsymbol{A}
 \end{gathered}$
 & $\begin{gathered}
 \boldsymbol{L}_M=-\frac{1}{c}\int d^3x
 \pi^{\mu}(\boldsymbol{x}\times\boldsymbol{\nabla})A_{\mu}
 \end{gathered}$ \\
\hline
{\centering Gauge-invariant decomposition\par}
 & $\begin{gathered}
 \boldsymbol{S}_D=\int d^3x \psi^{\dagger}\hat{\boldsymbol{\Sigma}}\psi
 \end{gathered}$
 & $\begin{gathered}
 \boldsymbol{L}^{\rm obs}_D=\int d^3x
 \psi^{\dagger}(\boldsymbol{x}\times\boldsymbol{p})\psi+\boldsymbol{L}_{\rm pure}
 \end{gathered}$
 & $\begin{gathered}
 \boldsymbol{S}^{\rm obs}_M=\varepsilon_0 \int d^3x\, \boldsymbol{E}_{\perp}\times\boldsymbol{A}_{\perp}
 \end{gathered}$
 & $\begin{gathered}
 \boldsymbol{L}^{\rm obs}_M=\varepsilon_0 \int d^3x\,E_{\perp}^{j} (\boldsymbol{x}\times\boldsymbol{\nabla})A_{\perp}^{j}
 \end{gathered}$ \\
\hline
{\centering Commutation relations\par}
 & $\begin{gathered}
 [S_{D,i},S_{D,j}] =i\hbar\varepsilon_{ijk}S_{D,k}
 \end{gathered}$
 & $\begin{gathered}
 [L^{\rm obs}_{D,i},L^{\rm obs}_{D,j}] =i\hbar\varepsilon_{ijk}L^{\rm obs}_{D,k}
 \end{gathered}$
 & $\begin{gathered}
 [S^{\rm obs}_{M,i},S^{\rm obs}_{M,j}]=0
 \end{gathered}$
 & {\color{red}$\begin{gathered}
 [L^{\rm obs}_{M,i},L^{\rm obs}_{M,j}] =i\hbar\varepsilon_{ijk}(L^{\rm obs}_{M,k}-S^{\rm obs}_{M,k})
 \end{gathered}$}\\[5pt]
\cline{2-5}
 & \multicolumn{2}{c|}{$[S_{D,i},L^{\rm obs}_{D,j}]=0$}
 & \multicolumn{2}{c|}{\textcolor{red}{$[L^{\rm obs}_{M,i},S^{\rm obs}_{M,j}]=i\hbar\varepsilon_{ijk}S^{\rm obs}_{M,k}$}}\\[5pt]
\hline
{\centering Total observable angular momentum\par}
 & \multicolumn{2}{c|}{$\begin{gathered}
 \boldsymbol{J}^{\rm obs}_D=\boldsymbol{S}_D+\boldsymbol{L}^{\rm obs}_D,\\
 [J^{\rm obs}_{D,i},J^{\rm obs}_{D,j}]=i\hbar\varepsilon_{ijk}J^{\rm obs}_{D,k}
 \end{gathered}$}
 & \multicolumn{2}{c|}{$\begin{gathered}
 \boldsymbol{J}^{\rm obs}_M=\boldsymbol{S}^{\rm obs}_M+\boldsymbol{L}^{\rm obs}_M,\\
 [J^{\rm obs}_{M,i},J^{\rm obs}_{M,j}]=i\hbar\varepsilon_{ijk}J^{\rm obs}_{M,k}
 \end{gathered}$} \\
\hline
\end{tabular}
\caption{\label{tab:gauge}
\textbf{Perfect vs. broken symmetry in angular momentum commutation relations of the photon:} In the canonical decomposition $\boldsymbol{J}=\boldsymbol{S}_{D}+\boldsymbol{L}_{D}+\boldsymbol{S}_{M}+\boldsymbol{L}_{M}$, the four angular momenta commute with each other and all of them satisfy the angular momentum commutation relations. But three of them are not gauge invariant. In $\boldsymbol{J}= \boldsymbol{S}_{D} + \boldsymbol{L}^{\rm obs}_{D}+\boldsymbol{S}^{\rm obs}_{M}+\boldsymbol{L}^{\rm obs}_{M}$ decomposition, all the four parts are gauge invariant. {\color{red}The photon spin and orbital contributions do not separately obey the standard angular-momentum algebra, whereas their sum does.}  The term $\boldsymbol{L}_{\rm pure}$ denotes the pure gauge contribution of the light to the Dirac orbital angular momentum (OAM), which will disappear in the Coulomb gauge.}
\end{table*}

\section{Gauge invariant decomposition of QED angular momentum}

We now put forth a gauge invariant theoretical framework to make connections to experimentally observable OAM and SAM. We note that photons are massless gauge bosons under the local $U(1)$-gauge symmetry of the standard (subscript ``ST") QED Lagrangian density $\mathcal{L}_{\rm QED, ST}=i\hbar c\bar{\psi}\gamma^{\mu}\partial_{\mu}\psi-mc^{2}\bar{\psi}\psi -qc\bar{\psi}\gamma_{\mu}A^{\mu}\psi +\mathcal{L}_{M,{\rm ST}}$. Thus, the gauge-dependence problem can only be fully solved by a theoretical framework that combines Dirac-Maxwell fields. This line of exploration is a paradigm shift from previous approaches in the field of photonics that do not use a fundamental QED theory including Dirac particles for addressing this problem. We argue that any measurement process of photon's SAM and OAM necessarily requires interaction with matter i.e. Dirac-Maxwell fields have to be analyzed as opposed to Maxwell fields alone. Thus, conservation laws which emerge from the combined Dirac-Maxwell-field angular momenta provides the clear path towards analyzing experimental observables. Schematically, this is depicted in Fig.~\ref{fig1} which deals with a relativistic quantum scattering experiment of a photon with a Dirac particle~\cite{drechsel2003dispersion}. We put forth the OAM and SAM conservation laws in this set-up to develop a theoretical framework for gauge invariant SAM and OAM observables.  We note that our definition of transverse and longitudinal follows the QED literature. It is unrelated to classical transverse concept defined in relation to the propagation direction for plane waves. 

We first start from a gauge non-invariant QED Lagrangian density 
\begin{equation}
\mathcal{L}_{\rm QED}=i\hbar c\bar{\psi}\gamma^{\mu}\partial_{\mu}\psi-mc^{2}\bar{\psi}\psi -qc\bar{\psi}\gamma_{\mu}A^{\mu}\psi +\mathcal{L}_M,
\end{equation}
where the gauge non-invariance arises from $\mathcal{L}_M$. We exploit a novel re-decomposition of the total angular momentum and enforcement of the Lorenz gauge condition to obtain the gauge-invariant SAM and OAM. To prove that our procedure is exact, we arrive at the same striking result through an alternative path where the gauge issue can be solved by quantizing the standard gauge-invariant Lagrangian density $\mathcal{L}_{\rm QED, ST}$ in the Coulomb gauge (see Appendix~\ref{sec:appendixF}).

The total angular momentum of the combined Dirac-Maxwell fields can be decomposed in two different ways which we term as canonical and gauge-invarant decomposition.\\
\textit{Canonical decomposition}: According to Noether's theorem, the total QED angular momentum obtained from $\mathcal{L}_{\rm QED}$ contains four parts $\boldsymbol{J}=\boldsymbol{S}_{D}+\boldsymbol{L}_{D}+\boldsymbol{S}_{M}+\boldsymbol{L}_{M}$. The SAM and OAM of the photon have been given in the previous part of the work. The SAM and OAM of the Dirac field are given by $\boldsymbol{S}_{D}=\frac{1}{2}\hbar\int d^{3}x\psi^{\dagger}\hat{\boldsymbol{\Sigma}}\psi$ and $\boldsymbol{L}_{D}=-i\hbar\int d^{3}x\psi^{\dagger}\boldsymbol{x}\times\boldsymbol{\nabla}\psi$, respectively. All four parts in the canonical decomposition satisfy the angular momentum commutation relations and they commute with each other. However, except for the Dirac spin, all the other three parts in $\boldsymbol{J}$ are not gauge invariant.

\textit{Gauge-invariant decomposition}: To obtain the gauge-invariant variables, we introduce the concept of gauge flow. Here, we extract the parts in $\boldsymbol{S}_{M}$ and $\boldsymbol{L}_{M}$ containing  scalar and longitudinal photons and flow them into the OAM of the Dirac field $\boldsymbol{L}_{D}$. Then, we obtain the gauge-invariant decomposition of the total angular momentum (superscript ``obs"):
\begin{equation}
\boldsymbol{S}_{D} + \boldsymbol{L}_{D}+\boldsymbol{S}_{M}+\boldsymbol{L}_{M}= \boldsymbol{J}= \boldsymbol{S}_{D} + \boldsymbol{L}^{\rm obs}_{D}+\boldsymbol{S}^{\rm obs}_{M}+\boldsymbol{L}^{\rm obs}_{M}.  \label{eq:gauge-hiding}
\end{equation}
In Tab.~\ref{tab:gauge}, we contrast the canonical decomposition of the total QED angular momentum with this gauge invariant decomposition. The gauge invariant part of our defined photon SAM and OAM operators recovers the angular momentum of light from CED~\cite{cohen1997photons}
$\boldsymbol{S}^{\rm obs}_{M}= \int d^{3}k \boldsymbol{s}_{\boldsymbol{k},3} =\varepsilon_{0}\!\int\! d^{3}x\boldsymbol{E}_{\perp}\!\times\!\boldsymbol{A}_{\perp}$
and
$\boldsymbol{L}^{\rm obs}_{M}\! =\! -i\hbar\!\!\int\!\! d^{3}k\!\!\sum_{\lambda=1,2}a_{\boldsymbol{k},\lambda}^{\dagger}(\boldsymbol{k}\times\boldsymbol{\nabla}_{\boldsymbol{k}})a_{\boldsymbol{k},\lambda}\!=\!\varepsilon_{0}\!\!\int\!\! d^{3}x E_{\perp}^{j}\boldsymbol{x}\times\boldsymbol{\nabla}A_{\perp}^{j}$. 
Here, $\boldsymbol{s}_{\boldsymbol{k},3}=i\hbar(a_{\boldsymbol{k},2}^{\dagger}a_{\boldsymbol{k},1}\!-\!a_{\boldsymbol{k},1}^{\dagger}a_{\boldsymbol{k},2})\boldsymbol{\epsilon}(\boldsymbol{k},3)$ is the observable spin density in $\boldsymbol{k}$-space and we have used the relation between $\boldsymbol{\pi}$ and the electric field $\boldsymbol{E} =-c\left(\partial^{0}\boldsymbol{A}+\boldsymbol{\nabla}A^{0}\right)=c\mu_{0}\boldsymbol{\pi}-c\boldsymbol{\nabla}A^{0}$. This shows that the gauge-invariant part of the photon spin $\boldsymbol{S}^{\rm obs}_M$ only contains information from the propagation direction within the full plane wave expansion [see Eq.~(\ref{eq:SM_matrix})]. We note that the $\boldsymbol{S}^{\rm obs}_M$ is not the total photon spin operator ~\cite{cohen1997photons} as it does not obey angular momentum commutation rules. Our result clearly shows that it is only the transverse-field sector of the total photon spin ($S_M$). 

The hallmark of our work is that our proposed Maxwellian SAM and OAM simultaneously recovers the correct OAM and SAM of the Dirac field. The gauge-invariant OAM of the Dirac field obtained from the above analysis is
\begin{equation}
\boldsymbol{L}^{\rm obs}_{D}=\int d^{3}x\psi^{\dagger}\boldsymbol{x}\times(-i\hbar\boldsymbol{\nabla})\psi + \boldsymbol{L}_{\rm pure},\label{eq:LDp}
\end{equation}
where $\boldsymbol{L}_{\rm pure}$ is the pure gauge contribution from the EM field (see Appendix~\ref{sec:appendixD}). We have verified that the mean value of $\boldsymbol{L}^{\rm obs}_{D}$ is gauge invariant.

There is another important physical observable related to circularly polarized photons and closely related to the photon spin---the photon helicity. Helicity is the magnitude of spin projection on the propagating direction of the particle, which is a Lorentz invariant scalar. Because only $\boldsymbol{S}^{\rm obs}_M$ has components in $\boldsymbol{k}$-direction, thus the photon helicity is given by
\begin{align}
\Lambda_M & = \int d^3 k \frac{\boldsymbol{s}_{\boldsymbol{k},3}\cdot\boldsymbol{k}}{|\boldsymbol{k}|} = i\hbar \int d^{3}k(a_{\boldsymbol{k},2}^{\dagger}a_{\boldsymbol{k},1}\!-\!a_{\boldsymbol{k},1}^{\dagger}a_{\boldsymbol{k},2}).
\end{align}

We argue that our decomposition  $\boldsymbol{J}=\boldsymbol{S}_{D}+\boldsymbol{L}^{\rm obs}_{D}+\boldsymbol{S}^{\rm obs}_{M}+\boldsymbol{L}^{\rm obs}_{M}$ embodies the correct physical behavior of QED angular momentum. We arrive at two new fundamental commutation relations for OAM of the Dirac fields as well as Maxwell fields
\begin{align}
[L^{\rm obs}_{D,i},L^{\rm obs}_{D,j}] & =i\hbar\varepsilon_{ijk}L_{D,k}^{\rm obs},\\
[L_{M,i}^{\rm obs},L_{M,j}^{\rm obs}] & =i\hbar\varepsilon_{ijk}L_{M,k}^{\rm obs}.
\end{align}
We note the perfect symmetry once again proving that experimentally observable OAM for Dirac-Maxwell fields follows the uncertainty principle. 

The definition of the Dirac spin is in agreement with previous literature, which certainly satisfies the standard angular momentum commutation relation $[S_{D,i},S_{D,j}]=i\hbar\varepsilon_{ijk}S_{D,k}$. However, the observable (gauge-invariant) photon spin operators do not obey the above mentioned symmetry of the full photon spin operator. As shown previously~\cite{van1994commutation,van1994spin}, the components of the transverse-field photon spin commute with each other 
\begin{equation}
[S^{\rm obs}_{M,i},S^{\rm obs}_{M,j}] = 0.
\end{equation}
The fundamental reason is because this latter observable spin operator does not contain the contribution from longitudinally polarized photons i.e. virtual photons. We emphasize that the full spin operators for Dirac-Maxwell fields obtained by us exhibits the perfect symmetry (Fig. 1).


\begin{table*}[t]
\centering
\scriptsize
\setlength{\tabcolsep}{3pt}
\renewcommand{\arraystretch}{1.45}
\begin{tabular}{|p{62pt}|p{56pt}|p{88pt}|p{62pt}|p{90pt}|p{50pt}|}
\hline
 & Dirac SAM & Dirac OAM & Maxwell SAM & Maxwell OAM & Quantized independent observables \\
\hline
Our decomposition
& $\begin{gathered}\frac{\hbar}{2}\int d^3x\,\psi^\dagger\hat{\boldsymbol\Sigma}\psi\end{gathered}$
& $\begin{gathered}\int d^3x\,\psi^\dagger (\boldsymbol x\times\boldsymbol p)\psi {}+\boldsymbol L_{\rm pure}\end{gathered}$
& $\begin{gathered}\varepsilon_0\int d^3x \boldsymbol E_\perp\times\boldsymbol A_\perp\end{gathered}$
& $\begin{gathered}\varepsilon_0\int d^3x\,E_\perp^j(\boldsymbol x\times\boldsymbol\nabla)A_\perp^j\end{gathered}$
& \textcolor{red}{No} \\
\hline
Belinfante~\cite{belinfante1939spin} decomposition
& \multicolumn{2}{c|}{$\begin{gathered}\boldsymbol J_D=\int d^3x\,\bar\psi\bigl[\boldsymbol x\times\tfrac12(\gamma^0i\boldsymbol D+\boldsymbol\gamma iD^0)\bigr]\psi\end{gathered}$}
& \multicolumn{2}{c|}{$\begin{gathered}\boldsymbol J_M=\varepsilon_0\int d^3x\boldsymbol x\times(\boldsymbol E\times\boldsymbol B)\end{gathered}$}
& No \\
\hline
Ji~\cite{Ji1997gauge} decomposition
& $\begin{gathered}\frac{\hbar}{2}\int d^3x\,\psi^\dagger\hat{\boldsymbol\Sigma}\psi\end{gathered}$
& $\begin{gathered}\frac{\hbar}{2}\int d^3x\,\psi^\dagger(\boldsymbol x\times i\boldsymbol D)\psi\end{gathered}$
& \multicolumn{2}{c|}{$\begin{gathered}\boldsymbol J_M=\varepsilon_0\int d^3x\boldsymbol x\times(\boldsymbol E\times\boldsymbol B)\end{gathered}$}
& No \\
\hline
Jaffe--Manohar~\cite{jaffe1990g1} decomposition
& $\begin{gathered}\frac{\hbar}{2}\int d^3x\,\psi^\dagger\hat{\boldsymbol\Sigma}\psi\end{gathered}$
& $\begin{gathered}\int d^3x\,\psi^\dagger(\boldsymbol x\times\boldsymbol p)\psi\end{gathered}$
& $\begin{gathered}\varepsilon_0\int d^3x\boldsymbol E\times\boldsymbol A\end{gathered}$
& $\begin{gathered}\varepsilon_0\int d^3x\,E^j(\boldsymbol x\times\boldsymbol\nabla)A^j\end{gathered}$
& No \\
\hline
Chen et al.~\cite{chen2008spin} decomposition
& $\begin{gathered}\frac{\hbar}{2}\int d^3x\,\psi^\dagger\hat{\boldsymbol\Sigma}\psi\end{gathered}$
& $\begin{gathered}\int d^3x\,\psi^\dagger\boldsymbol x\times(\boldsymbol p-q\boldsymbol A_\parallel)\psi\end{gathered}$
& $\begin{gathered}\varepsilon_0\int d^3x\boldsymbol E\times\boldsymbol A_\perp\end{gathered}$
& $\begin{gathered}\varepsilon_0\int d^3x\,E^j(\boldsymbol x\times\boldsymbol\nabla)A_\perp^j\end{gathered}$
& No \\
\hline
Wakamatsu~\cite{Wakamatsu2010gauge} decomposition
& $\begin{gathered}\frac{\hbar}{2}\int d^3x\,\psi^\dagger\hat{\boldsymbol\Sigma}\psi\end{gathered}$
& $\begin{gathered}\int d^3x\,\psi^\dagger \boldsymbol x\times(\boldsymbol p-q\boldsymbol A)\psi\end{gathered}$
& $\begin{gathered}\varepsilon_0\int d^3x \boldsymbol E\times\boldsymbol A_\perp\end{gathered}$
& $\begin{gathered}\varepsilon_0\int d^3x\,\bigl[ E^j(\boldsymbol x\times\boldsymbol\nabla)A_\perp^j \\ +(\boldsymbol\nabla\cdot\boldsymbol E) (\boldsymbol x\times\boldsymbol\nabla)\boldsymbol A_\perp\bigr]\end{gathered}$
& No \\
\hline
\end{tabular}
\caption{\label{tab:3}We contrast our gauge-invariant decomposition of the QED angular momentum with previous important results that have inspired us to discover new operators (please refer to Appendix~\ref{sec:appendixG} for more details). Our work includes the role of virtual photons which is a fundamental tenet for observing symmetry between Dirac and Maxwell fields. This goes beyond the traditional QED Lagrangian density used in previous work. Furthermore, \textcolor{red}{the four terms in our gauge-invariant decomposition represent separately defined contributions, although their Cartesian components do not all commute with each other. In particular, the observable photon SAM and OAM satisfy the nonvanishing spin--orbital commutation relation, while their sum obeys the standard angular-momentum algebra} (see Appendix~\ref{sec:appendixE}), thus they can be measured independently. Here, $D_{\mu} =\partial_{\mu} + iq A_{\mu}$ is the covariant derivative.}
\end{table*}

As shown in Table~\ref{tab:3}, our work marks a departure from previous decompositions~\cite{jaffe1990g1,Ji1997gauge,chen2008spin,Wakamatsu2010gauge}. The four parts in the gauge-invariant decomposition of $\boldsymbol{J}$ [right-hand side of Eq.~(\ref{eq:gauge-hiding})] represent separately defined observable contributions, although their Cartesian components do not all commute with each other. Specifically, \textcolor{red}{as previously established by van Enk and Nienhuis~\cite{van1994commutation,van1994spin} and reproduced in Appendix~\ref{sec:appendixE}, the observable photon spin and OAM satisfy the nonvanishing cross-commutation relation
$[L^{\rm obs}_{M,i},S^{\rm obs}_{M,j}]
=i\hbar\varepsilon_{ijk}S^{\rm obs}_{M,k}$. We note that the (gauge-invariant) observable total photon angular momentum $\boldsymbol{J}^{\rm obs}_M = \boldsymbol{L}^{\rm obs}_M + \boldsymbol{S}^{\rm obs}_M$ generates spatial rotations on the physical photon Hilbert space, whereas the full canonical angular momentum $\boldsymbol{J}_M = \boldsymbol{L}_M + \boldsymbol{S}_M$ acts on the covariantly quantized field space. Consistent with the known spin--orbital algebra, their observable sum recovers the standard angular momentum commutation relation,
$[J^{\rm obs}_{M,i},J^{\rm obs}_{M,j}]
=i\hbar\varepsilon_{ijk}J^{\rm obs}_{M,k}$.
}

Recently the photonic spin density has been directly measured through interaction of light with a room temperature quantum magnetometer~\cite{Farid2021quantum}. Nitrogen vacancy (NV) centers in diamond function as spin qubits which are sensitive to magnetic fields as well as magnetic field fluctuations. It was shown that a detuned laser beam which is circularly polarized creates energy level shifts in the ground spin states of an NV center that are analogous to an effective static magnetic field. The experiment shows a coherent interaction between the local spin density vector and the NV center which is read out using the Ramsey interference protocol. We note that the specific physical quantity which is being measured in this experiment is the observable photonic spin density vector ($\boldsymbol{S}^{\rm obs}_{M}$). One future possibility is for this same experiment to be extended to vector magnetometry  using an OAM beam or pulse.  In the plane perpendicular to the propagating axis, the mean values of both the photon spin and OAM vanish. However, their quantum uncertainties are not zero~\cite{Yang2021nonclassical}. Thus, the Heisenberg's uncertainty relations for photonic OAM operators can be verified by measuring their quantum fluctuations in orthogonal directions. This vector measurement is possible through the relative alignment of the NV axis and the incident light beam. Low temperatures will be needed to extend the coherence time of the NV center to increase the sensitivity and measure both mean values and fluctuations of the observable photonic spin density vector. Therefore local photonic spin density and spin noise can be measured via a nitrogen-vacancy (NV) center, which functions as a nano-scale photonic spin density sensor~\cite{Farid2021quantum}. Along similar lines, the orbital photogalvanic effect~\cite{ji2020photocurrent,lai2022direct} could be exploited to measure the photonic OAM density and quantum fluctuations of OAM in the transverse plane~\cite{Yang2021nonclassical}.

\section{Conclusion} 
We have discovered the quantum operator of the photon spin providing the first QED theory of angular momentum. Our approach involves Dirac-Maxwell fields in a quantum gauge theory framework. We have proven a perfect symmetry between the Dirac and Maxwell quantum spin operator commutation relations. In experiment, our theory can be verified through interaction of photonic spin density with an NV center~\cite{Farid2021quantum} and the interaction of OAM density with two-dimensional materials~\cite{ji2020photocurrent,lai2022direct}.


This work is supported by the funding from DARPA Nascent Light-Matter Interactions. L.P.Y has also been funded by National Key R\&D Program of China (Grant No. 2021YFE0193500).

\appendix

\section{Angular momentum of light from Noether's theorem\label{sec:appendixA}}
To obtain enough polarization degrees of freedom to construct the full photon spin operator, we use the following Maxwell Lagrangian density to quantize the electromagnetic (EM) field in the Lorenz gauge covariantly~\cite{cohen1997photons,greiner2013field},
\begin{equation}
\mathcal{L}_{M}=-\frac{1}{2\mu_{0}}(\partial_{\mu}A^{\nu})(\partial^{\mu}A_{\nu}).
\end{equation}
We note that, different from the standard EM Lagrangian density $\mathcal{L}_{M,{\rm ST}} = -(1/4\mu_0) F^{\mu\nu}F_{\mu\nu}$,  $\mathcal{L}_M$ itself is not gauge invariant. We will show how to eliminate the gauge dependence in the physical quantities via the gauge condition in the following.

These two Lagrangian density $\mathcal{L}_M$ and $\mathcal{L}_{M,{\rm ST}}$ can be connected via~\cite{greiner2013field,cohen1997photons}
\begin{equation}
\mathcal{L}_M =  \mathcal{L}_{M,{\rm ST}} -\frac{1}{2} (\partial_{\mu}A^{\mu})^2+\frac{1}{2} \partial_{\mu}[A_{\nu}(\partial^{\nu}A^{\mu})-(\partial_{\nu}A^{\nu})A^{\mu}]. 
\end{equation}
The last four divergence term can be dropped since the EM fields are assumed to tend to 0 sufficiently fast at infinity. Under the Lorenz gauge condition $\partial_{\mu}A^{\mu}=0$, these two Lagrangian densities are equivalent to each other.

Noether's theorem tells us that if the action $W=\int d^4x \mathcal{L}$  keeps invariant under a continuous symmetry transformation, a conserved quantity can be obtained (see Chap.2 in Ref.~\cite{greiner2013field}).
Applying the Noether's theorem to the translation symmetry, we obtain the canonical energy-momentum tensor
\begin{align}
\Theta_{M}^{\mu\nu} & =\frac{\partial\mathcal{L}_{M}}{\partial(\partial_{\mu}A_{\sigma})}\partial^{\nu}A_{\sigma}-g^{\mu\nu}\mathcal{L}_{M}\\
& =-\frac{1}{\mu_{0}}(\partial^{\mu}A^{\sigma})(\partial^{\nu}A_{\sigma})+\frac{1}{2\mu_{0}}g^{\mu\nu}(\partial^{\rho}A^{\sigma})(\partial_{\rho}A_{\sigma}).
\end{align}
This leads to the conserved four-momentum vector
\begin{align}
P^{\nu}_M\!=\!\Theta_{M}^{0,\nu}\! =\!-\frac{1}{\mu_{0}}(\partial^{0}A^{\sigma})(\partial^{\nu}A_{\sigma})\!+\!\frac{1}{2\mu_{0}}g^{0\nu}(\partial^{\rho}A^{\sigma})(\partial_{\rho}A_{\sigma}).
\end{align}
The time component of $P^{\nu}_M$ gives the Hamiltonian (energy) of the
system $H_M = P^{0}_M=\int d^{3}x\mathcal{H}_{M}$,
with Hamiltonian density
\begin{align}
\mathcal{H}_{M} & =-\frac{1}{2\mu_{0}}\left[(\partial^{0}A^{\sigma})(\partial^{0}A_{\sigma})+(\boldsymbol{\nabla}A^{\sigma})\cdot(\boldsymbol{\nabla}A_{\sigma})\right].
\end{align}
The spatial part of $P^{\nu}_M$ gives the momentum of the EM field
\begin{equation}
\boldsymbol{P}_{M}=\varepsilon_{0}\int d^{3}x\dot{A}^{\sigma}\boldsymbol{\nabla}A_{\sigma},
\end{equation}
where we have divided an extra constant $c$ to get the correct dimension.

Similarly, applying Noether's theorem to the three-dimensional rotation symmetry~\cite{greiner2013field}, we obtain the angular momentum tensor density 
\begin{align}
M_{M}^{\mu\nu\lambda} & =\Theta_{M}^{\mu\lambda}x^{\nu}-\Theta_{M}^{\mu\nu}x^{\lambda}+\frac{\partial\mathcal{L}_{M}}{\partial(\partial_{\mu}A^{\sigma})}(I^{\nu\lambda})^{\sigma\tau}A_{\tau}
\end{align}
where infinitesimal generator of the Lorentz group for the vector field is given by
\begin{equation}
(I^{\alpha\beta})^{\mu\nu}=g^{\alpha\mu}g^{\beta\nu}-g^{\alpha\nu}g^{\beta\mu}.
\end{equation}
This generator is a an antisymmetric matrix
\begin{equation}
(I^{\alpha\beta})^{\mu\nu}=-(I^{\beta\alpha})^{\mu\nu}.
\end{equation}
Utilizing $\Theta_{M}^{\mu\nu}$ and $\mathcal{L}_{M}$, 
we have
\begin{align}
M_{M}^{\mu\nu\lambda} \!& =\!\Theta_{M}^{\mu\lambda}x^{\nu}\!-\!\Theta_{M}^{\mu\nu}x^{\lambda}\!-\!\frac{1}{\mu_{0}}(\partial^{\mu}A_{\sigma})(g^{\nu\sigma}g^{\lambda\tau}\!-\!g^{\nu\tau}g^{\lambda\sigma})A_{\tau}\\
 & =\Theta_M^{\mu\lambda}x^{\nu}-\Theta_M^{\mu\nu}x^{\lambda}-\frac{1}{\mu_{0}}[(\partial^{\mu}A^{\nu})A^{\lambda}-(\partial^{\mu}A^{\lambda})A^{\nu}].
\end{align}

The total angular momentum $M_M^{ij}$ can be obtained from the angular momentum density $M_M^{\mu\nu\lambda}$ by setting $\mu=0$ and taking spatial components of $\nu$ and $\lambda$. Here, we split the total angular momentum into two parts. The first one denotes the OAM of light $
L_{M}^{ij} =\int d^{3}x(\Theta_{M}^{0j}x^{i}-\Theta_{M}^{0i}x^{j})$ and the second part describes the "intrinsic" angular momentum (the spin) of the vector field
$S_{M}^{ij} =-\frac{1}{\mu_{0}}\int d^{3}x[(\partial^{0}A^{i})A^{j}-(\partial^{0}A^{j})A^{i}]$, with $i,j=x,y,z$. It is straightforward to see that, in terms of
three-vectors with relation~$J_{M}^{k}=\frac{1}{2}\epsilon^{ijk}M^{ij}$~\cite{greiner2013field},
the OAM and SAM of light can be rewritten as
\begin{align}
\boldsymbol{L}_{M} & =\frac{1}{\mu_{0}c}\int d^{3}x(\partial_{0}A^{\mu})\boldsymbol{x}\times\boldsymbol{\nabla}A_{\mu},\\
\boldsymbol{S}_{M} & =-\frac{1}{\mu_{0}c}\int d^{3}x[(\partial_{0}\boldsymbol{A})\times\boldsymbol{A}].
\end{align}

\section{Quantum commutation relations for angular momenta\label{sec:appendixB}} 
We now show how to obtain the standard quantum commutation relations for the angular momentum of light within the canonical quantization framework. We first define the canonical conjugate momentum of light
\begin{equation}
\pi_{\mu}=\frac{\partial\mathcal{L}_{M}}{\partial(\partial_{0}A^{\mu})}=-\frac{1}{\mu_{0}}\partial^{0}A_{\mu}. \label{eq:pi_mu}
\end{equation}
Then, we can re-write the Hamiltonian, SAM, and OAM of the photon as
\begin{align}
H_M & = -\frac{1}{2\mu_{0}}\int d^{3}x\left[\mu_0^2\pi^{\mu}\pi_{\mu}+(\boldsymbol{\nabla}A^{\mu})\cdot(\boldsymbol{\nabla}A_{\mu})\right],\\
\boldsymbol{S}_{M} & =\frac{1}{c}\int d^{3}x\boldsymbol{\pi}\times\boldsymbol{A},\\
\boldsymbol{L}_{M} & = -\frac{1}{c}\int d^{3}x\pi^{\mu}\boldsymbol{x}\times\boldsymbol{\nabla}A_{\mu}
\end{align}

The EM field is quantized by postulating the equal-time commutation relations,
\begin{align}
[A^{\mu}(\boldsymbol{x},t),\pi^{\nu}(\boldsymbol{x}',t)] & =i\hbar cg^{\mu\nu}\delta^{3}(\boldsymbol{x}-\boldsymbol{x}'), \label{eq:ETCR-M1}\\
[A^{\mu}(\boldsymbol{x},t),A^{\nu}(\boldsymbol{x}',t)] & =[\pi^{\mu}(\boldsymbol{x},t),\pi^{\nu}(\boldsymbol{x}',t)]=0.\label{eq:ETCR-M2}
\end{align}
We now verify the canonical angular momentum commutation relations\begin{widetext}
\begin{align}
[L_{M,i},L_{M,j}]\nonumber = & \frac{1}{c^{2}}\int d^{3}x\int d^{3}x'[\pi_{m}(\boldsymbol{x})(\boldsymbol{x}\times\boldsymbol{\nabla})_{i}A_{m}(\boldsymbol{x}),\pi_{m'}(\boldsymbol{x}')(\boldsymbol{x}'\times\boldsymbol{\nabla}')_{j}A_{m'}(\boldsymbol{x}')]\\
= & -\frac{i\hbar}{c}\!\int\!\! d^{3}x\!\!\int\!\! d^{3}x'\left\{ \pi_{m}(\boldsymbol{x}')(\boldsymbol{x}'\times\boldsymbol{\nabla}')_{j}\delta^{3}(\boldsymbol{x}-\boldsymbol{x}')(\boldsymbol{x}\times\boldsymbol{\nabla})_{i}A_{m}(\boldsymbol{x})-\pi_{m}(\boldsymbol{x})(\boldsymbol{x}\times\boldsymbol{\nabla})_{i}\delta^{3}(\boldsymbol{x}-\boldsymbol{x}')(\boldsymbol{x}'\times\boldsymbol{\nabla}')_{j}A_{m}(\boldsymbol{x}')\right\} \\
= & -\frac{i\hbar}{c}\int d^{3}x\left\{ \pi_{m}(\boldsymbol{x})(\boldsymbol{x}\times\boldsymbol{\nabla})_{j}(\boldsymbol{x}\times\boldsymbol{\nabla})_{i}A_{m}(\boldsymbol{x})-\pi_{m}(\boldsymbol{x})(\boldsymbol{x}\times\boldsymbol{\nabla})_{i}(\boldsymbol{x}\times\boldsymbol{\nabla})_{j}A_{m}(\boldsymbol{x})\right\} \\
= & -\frac{i\hbar}{c}\int d^{3}x\left\{ \pi_{m}(\boldsymbol{x})\varepsilon_{ijk}(\boldsymbol{x}\times\boldsymbol{\nabla})_{k}A_{m}(\boldsymbol{x})\right\} =i\hbar\varepsilon_{ijk}L_{M,k},
\end{align}
\begin{align}
[S_{M,i},S_{M,j}]= & \frac{1}{c^{2}}\varepsilon_{ikl}\varepsilon_{jk'l'}\int d^{3}x\int d^{3}x'[\pi_{k}(\boldsymbol{x})A_{l}(\boldsymbol{x}),\pi_{k'}(\boldsymbol{x}')A_{l'}(\boldsymbol{x}')]\\
= & \frac{1}{c^{2}}\varepsilon_{ikl}\varepsilon_{jk'l'}\int d^{3}x\int d^{3}x'\left\{ \pi_{k}(\boldsymbol{x})[A_{l}(\boldsymbol{x}),\pi_{k'}(\boldsymbol{x}')]A_{l'}(\boldsymbol{x}')+\pi_{k'}(\boldsymbol{x}')[\pi_{k}(\boldsymbol{x}),A_{l'}(\boldsymbol{x}')]A_{l}(\boldsymbol{x})\right\} \\
= & -\frac{i\hbar}{c}d^{3}x\left[\varepsilon_{ikl}\varepsilon_{jll'}\pi_{k}(\boldsymbol{x})A_{l'}(\boldsymbol{x})-\varepsilon_{ikl}\varepsilon_{jk'k}\pi_{k'}(\boldsymbol{x})A_{l}(\boldsymbol{x})\right]=i\hbar\varepsilon_{ijk}S_{M,k},
\end{align}
\begin{align}
[L_{M,i},S_{M,j}] = & -\frac{1}{c^{2}}\int d^{3}x\int d^{3}x'[\pi_{m}(\boldsymbol{x})(\boldsymbol{x}\times\boldsymbol{\nabla})_{i}A_{m}(\boldsymbol{x}),\varepsilon_{jkl}\pi_{k}(\boldsymbol{x}')A_{l}(\boldsymbol{x}')]\\
= & \frac{i\hbar}{c}\varepsilon_{0}\int d^{3}x\int d^{3}x'\varepsilon_{jkl}\left\{ \pi_{k}(\boldsymbol{x})(\boldsymbol{x}\times\boldsymbol{\nabla})_{i}\delta^{3}(\boldsymbol{x}-\boldsymbol{x}')A_{l}(\boldsymbol{x}')-\pi_{k}(\boldsymbol{x}')\delta^{3}(\boldsymbol{x}-\boldsymbol{x}')(\boldsymbol{x}\times\boldsymbol{\nabla})_{i}A_{l}(\boldsymbol{x})\right\} \\
= & -\frac{i\hbar}{c}\varepsilon_{0}\int d^{3}x\int d^{3}x'\varepsilon_{jkl}\left\{ \left[(\boldsymbol{x}\times\boldsymbol{\nabla})_{i}\pi_{k}(\boldsymbol{x})\right]A_{l}(\boldsymbol{x})+\pi_{k}(\boldsymbol{x})(\boldsymbol{x}\times\boldsymbol{\nabla})_{i}A_{l}(\boldsymbol{x})\right\} \\
= & -\frac{i\hbar}{c}\varepsilon_{0}\int d^{3}x\int d^{3}x'\varepsilon_{jkl}\left\{ -\pi_{k}(\boldsymbol{x})(\boldsymbol{x}\times\boldsymbol{\nabla})_{i}A_{l}(\boldsymbol{x})+\pi_{k}(\boldsymbol{x})(\boldsymbol{x}\times\boldsymbol{\nabla})_{i}A_{l}(\boldsymbol{x})\right\} =0,
\end{align}\end{widetext}
where we have used the partial integral techniques and the identities
\begin{equation}
\varepsilon_{ikm}\varepsilon_{jlm}=\delta_{ij}\delta_{kl}-\delta_{il}\delta_{km}.
\end{equation}

\section{Key role of longitudinal degrees of freedom \label{sec:plane-wave expansion}}
To clearly show the role of polarization degrees of the Maxwell fields, we perform the plane-wave expansions to the field operators as given in Eqs.~(\ref{eq:Amu}) and (\ref{eq:pimu}). The plane-wave expansion of the Dirac field has been well studied. Here, we mainly focus on the expansion of the SAM and OAM of photons. The polarization unit vectors in Eqs.~(\ref{eq:Amu}) and (\ref{eq:pimu}) satisfy the four-dimensional orthonormal conditions
\begin{equation}
\epsilon_{\mu}(\boldsymbol{k},\lambda)\epsilon^{\mu}(\boldsymbol{k},\lambda')=g_{\lambda\lambda'}.
\end{equation}
and the covariant completeness relation~\cite{greiner2013field}
\begin{equation}
    \sum_{\lambda=0}^3 g_{\lambda\lambda} \epsilon_{\mu}(\boldsymbol{k},\lambda)\epsilon_{\nu}(\boldsymbol{k},\lambda) = g_{\mu\nu},
\end{equation}
where $g_{\lambda\lambda}$ at the left hand side denotes the sign $\pm 1$ instead of the metric tensor. Here, we emphasize that the plane-wave expansion of $\pi^{\mu}$ is not obtained via the relation in Eq.~(\ref{eq:pi_mu}). Equations (\ref{eq:Amu}) and (\ref{eq:pimu}) function more like the definition of the ladder operators. We note that the photonic angular momentum relations in Eqs. (\ref{MCR1}-\ref{MCR3}) can also be verified \textcolor{red}{for the full canonical field with the complete polarization-frame connection retained} via their plane-wave expansions in the $\boldsymbol{k}$-space as shown in Appendix~\ref{sec:CR-recheck}.

We can re-express the ladder operators $a_{\boldsymbol{k},\lambda}$ and $a_{\boldsymbol{k},\lambda}^{\dagger}$ with canonical field variables $A^{\mu}$ and $\pi^{\mu}$ via the inverse transformation\begin{widetext}
\begin{align}
a_{\boldsymbol{k},\lambda} & = g_{\lambda\lambda}\!\!\int\!\! d^3x  \sqrt{\frac{\varepsilon_0}{2\hbar\omega_{\boldsymbol{k}}(2\pi)^3}} \left[\omega_{\boldsymbol{k}}A^\mu(\boldsymbol{x})-\frac{i}{\varepsilon_0}\pi^{\mu}(\boldsymbol{x})\right]\epsilon_{\mu}(\boldsymbol{k},\lambda) e^{-i\boldsymbol{k}\cdot\boldsymbol{x}},\\ 
a_{\boldsymbol{k},\lambda}^{\dagger} & = g_{\lambda\lambda} \int d^3x  \sqrt{\frac{\varepsilon_0}{2\hbar\omega_{\boldsymbol{k}}(2\pi)^3}} \left[\omega_{\boldsymbol{k}}A^\mu(\boldsymbol{x})+\frac{i}{\varepsilon_0}\pi^{\mu}(\boldsymbol{x})\right]\epsilon_{\mu}(\boldsymbol{k},\lambda) e^{i\boldsymbol{k}\cdot\boldsymbol{x}}.
\end{align}\end{widetext}
Utilizing the quantization ansatz (\ref{eq:ETCR-M1}) and (\ref{eq:ETCR-M2}) for the photon, we can verify that the ladder operators satisfy the bosonic commutations
\begin{align}
[a_{\boldsymbol{k},\lambda},a_{\boldsymbol{k}',\lambda'}^{\dagger}]& =-g_{\lambda\lambda'}\delta^{3}(\boldsymbol{k}-\boldsymbol{k}'), \label{eq:BCR1} \\
[a_{\boldsymbol{k},\lambda},a_{\boldsymbol{k}',\lambda'}]& =[a_{\boldsymbol{k},\lambda}^{\dagger},a_{\boldsymbol{k}',\lambda'}^{\dagger}]=0.\label{eq:BCR2}
\end{align}

The plane-wave expansion of the Hamiltonian $H_M$ and the momentum $\boldsymbol{P}_M$ of the photon have been given in the textbook~\cite{greiner2013field,cohen1997photons}. Here, we give some details,
\begin{align}
H_{M}  = & \int d^{3}x\mathcal{H}_{M}=-\frac{1}{2}\int\left(\mu_{0}\pi^{\mu}\pi_{\mu}+\frac{1}{\mu_{0}}(\boldsymbol{\nabla}A^{\mu})\cdot(\boldsymbol{\nabla}A_{\mu})\right)d^{3}x \nonumber\\
= & -\!\!\!\int\!\! d^{3}x\!\!\!\int\!\! d^{3}k\!\!\!\int\!\!d^{3}k'\!\sum_{\lambda\lambda'}\frac{\hbar}{4(2\pi)^{3}\sqrt{\omega_{\boldsymbol{k}}\omega_{\boldsymbol{k}'}}}\epsilon^{\mu}(\boldsymbol{k},\lambda)\epsilon_{\mu}(\boldsymbol{k}',\lambda')\nonumber\\ & \!\!\!\!\times\left(\omega_{\boldsymbol{k}}\omega_{\boldsymbol{k}'}\!+\!c^{2}\boldsymbol{k}\cdot\boldsymbol{k}'\right)\left[a_{\boldsymbol{k},\lambda}a_{\boldsymbol{k}',\lambda'}^{\dagger}e^{-i(k-k')\cdot x} \!+\!a_{\boldsymbol{k},\lambda}^{\dagger}a_{\boldsymbol{k}',\lambda'}e^{i(k-k')\cdot x}\right. \nonumber\\
& \left. -a_{\boldsymbol{k},\lambda}a_{\boldsymbol{k}',\lambda'}e^{-i(k+k')\cdot x}-a_{\boldsymbol{k},\lambda}^{\dagger}a_{\boldsymbol{k}',\lambda'}^{\dagger}e^{i(k+k')\cdot x}\right]\nonumber\\
= & -\!\!\int \!vd^{3}k\frac{\hbar\omega_{\boldsymbol{k}}}{2}\!\sum_{\lambda\lambda'}\epsilon^{\mu}(\boldsymbol{k},\lambda)\epsilon_{\mu}(\boldsymbol{k},\lambda')\left(a_{\boldsymbol{k},\lambda}a_{\boldsymbol{k},\lambda'}^{\dagger}\!+\!a_{\boldsymbol{k},\lambda}^{\dagger}a_{\boldsymbol{k},\lambda'}\right)\nonumber\\
= & -\frac{1}{2}\int d^{3}k\hbar\omega_{\boldsymbol{k}}\sum_{\lambda}g_{\lambda\lambda}\left(a_{\boldsymbol{k},\lambda}a_{\boldsymbol{k},\lambda}^{\dagger}+a_{\boldsymbol{k},\lambda}^{\dagger}a_{\boldsymbol{k},\lambda}\right)\\
=  & \!\!\int\!\! d^{3}k\hbar\omega_{\boldsymbol{k}}\left(a_{\boldsymbol{k},1}^{\dagger}a_{\boldsymbol{k},1}\!+\!a_{\boldsymbol{k},2}^{\dagger}a_{\boldsymbol{k},2}\!+\!a_{\boldsymbol{k},3}^{\dagger}a_{\boldsymbol{k},3}\!-\!a_{\boldsymbol{k},0}^{\dagger}a_{\boldsymbol{k},0}\right),
\end{align}
where the normal-ordering has been taken in the last step. We note that the counter-rotating wave terms (i.e., $a_{\boldsymbol{k},\lambda}a_{\boldsymbol{k}',\lambda'}$ and $a_{\boldsymbol{k},\lambda}^{\dagger}a^{\dagger}_{\boldsymbol{k}',\lambda'}$) vanish due to $(\omega_{\boldsymbol{k}}\omega_{\boldsymbol{k}'}+c^{2}\boldsymbol{k}\cdot\boldsymbol{k}')=0$ when $\boldsymbol{k}=-\boldsymbol{k}'$. No rotating wave approximation has been taken here. 

We also note that there are three main problems in the covariant quantization in the Lorenz gauge: (1) the Hamiltonian is not gauge invariant; (2) the frequency of the scalar photon is negative; (3) the norm of the scalar-photon state can be negative, i.e., $\langle 0|a_{\boldsymbol{k},0}a_{\boldsymbol{k},0}^{\dagger}|0\rangle=-1$. The first two problems can be solved simultaneously by enforcing the Gupta-Bleuler constraint ~\cite{gupta1950theory,bleuler1950new}, which is the quantum version of the Lorenz gauge condition. This gauge condition is essential to obtain the quantum Maxwell equation and to remove the gauge dependence in the Lorenz-gauge quantization framework. The last problem can be solved by exploiting Dirac's indefinite metric in space of quantum states~\cite{Pauli1943on,gupta1950theory,bleuler1950new} (also see Chap. V in the book~\cite{cohen1997photons}).

Similarly, the momentum of the EM field is expanded as
\begin{align}
\boldsymbol{P}_{M} & = \varepsilon_{0}\int d^{3}x\dot{A}^{\sigma}\boldsymbol{\nabla}A_{\sigma}=-\frac{1}{c}\int d^{3}x\pi^{\mu}\boldsymbol{\nabla}A_{\mu} \\
& \!=\!\! \int\!\! d^{3}k\left(a_{\boldsymbol{k},1}^{\dagger}a_{\boldsymbol{k},1}\!+\!a_{\boldsymbol{k},2}^{\dagger}a_{\boldsymbol{k},2}\!+\!a_{\boldsymbol{k},3}^{\dagger}a_{\boldsymbol{k},3}\!-\!a_{\boldsymbol{k},0}^{\dagger}a_{\boldsymbol{k},0}\right)\hbar\boldsymbol{k}.\label{eq:PM_planewave} 
\end{align}
In the plane-wave expansion of $\boldsymbol{P}_M$, fast-oscillating counter-rotating wave terms also cancel out with each other, i.e., 
\begin{align}
  &\frac{1}{2}\sum_{\lambda\lambda'}\int d^3k \hbar\boldsymbol{k}a_{\boldsymbol{k},\lambda}a_{-\boldsymbol{k},\lambda'}\epsilon^\mu(\boldsymbol{k},\lambda)\epsilon_\mu(-\boldsymbol{k},\lambda') \nonumber \\  =&\frac{1}{4}\sum_{\lambda\lambda'}\int d^3k \hbar\boldsymbol{k}(a_{\boldsymbol{k},\lambda}a_{-\boldsymbol{k},\lambda'}+a_{\boldsymbol{-k},\lambda'}a_{\boldsymbol{k},\lambda})\epsilon^\mu(\boldsymbol{k},\lambda)\epsilon_\mu(-\boldsymbol{k},\lambda')\nonumber\\
  = & \frac{1}{4}\sum_{\lambda\lambda'}\int d^3k \hbar\boldsymbol{k}a_{\boldsymbol{k},\lambda}a_{-\boldsymbol{k},\lambda'}[\epsilon^\mu(\boldsymbol{k},\lambda)\epsilon_\mu(-\boldsymbol{k},\lambda')\nonumber \\ & -\epsilon^\mu(-\boldsymbol{k},\lambda')\epsilon_\mu(\boldsymbol{k},\lambda)] =0, 
\end{align}
where we have used the property $[a_{\boldsymbol{k},\lambda},a_{-\boldsymbol{k},\lambda'}]=0$ and the four-vector inner product does not dependent on the order of the two vectors. Similar argument shows that $a^{\dagger}_{\boldsymbol{k}\lambda},a^{\dagger}_{-\boldsymbol{k},\lambda'}$ terms also vanish.

The SAM of the Maxwell field can be expanded as\begin{widetext}
\begin{align}
\boldsymbol{S}_{M}= & -\frac{1}{\mu_{0}c}\int d^{3}x(\partial_{0}\boldsymbol{A})\times\boldsymbol{A}=\frac{1}{c}\int d^{3}x\boldsymbol{\pi}\times\boldsymbol{A}\\
= & \frac{i\hbar}{2(2\pi)^{3}}\!\!\int\!\! d^{3}k\!\!\int\!\! d^{3}k'\!\!\int\!\! d^{3}x\!\sum_{\lambda,\lambda'=1}^{3}\sqrt{\frac{\omega_{\boldsymbol{k}}}{\omega_{\boldsymbol{k}'}}}\left[a_{\boldsymbol{k},\lambda}\boldsymbol{\epsilon}(\boldsymbol{k},\lambda)e^{i\boldsymbol{k}\cdot \boldsymbol{x}}\!-\!a_{\boldsymbol{k},\lambda}^{\dagger}\boldsymbol{\epsilon}(\boldsymbol{k},\lambda)e^{-i\boldsymbol{k}\cdot \boldsymbol{x}}\right]\left[a_{\boldsymbol{k}',\lambda'}\boldsymbol{\epsilon}(\boldsymbol{k}',\lambda')e^{i\boldsymbol{k'}\cdot\boldsymbol{x} }+a_{\boldsymbol{k}',\lambda'}^{\dagger}\boldsymbol{\epsilon}(\boldsymbol{k},\lambda')e^{-i\boldsymbol{k}'\cdot\boldsymbol{x} }\right]\\
= & i\hbar\!\!\int\!\! d^{3}k\left[\left(a_{\boldsymbol{k},3}^{\dagger}a_{\boldsymbol{k},2}\!-\!a_{\boldsymbol{k},2}^{\dagger}a_{\boldsymbol{k},3}\right)\boldsymbol{\epsilon}(\boldsymbol{k},1)\!+\!\left(a_{\boldsymbol{k},1}^{\dagger}a_{\boldsymbol{k},3}\!-\!a_{\boldsymbol{k},3}^{\dagger}a_{\boldsymbol{k},1}\right)\boldsymbol{\epsilon}(\boldsymbol{k},2)\!+\!\left(a_{\boldsymbol{k},2}^{\dagger}a_{\boldsymbol{k},1}\!-\!a_{\boldsymbol{k},1}^{\dagger}a_{\boldsymbol{k},2}\right)\boldsymbol{\epsilon}(\boldsymbol{k},3)\right]\equiv i\hbar\int d^{3}k\sum_{\lambda=1}^{3}\boldsymbol{s}_{\boldsymbol{k},\lambda},
\end{align}\end{widetext}
with
\begin{align}
\boldsymbol{s}_{\boldsymbol{k},1} & =(a_{\boldsymbol{k},3}^{\dagger}a_{\boldsymbol{k},2}-a_{\boldsymbol{k},2}^{\dagger}a_{\boldsymbol{k},3})\boldsymbol{\epsilon}(\boldsymbol{k},1),\label{eq:sk1}\\
\boldsymbol{s}_{\boldsymbol{k},2} & =(a_{\boldsymbol{k},1}^{\dagger}a_{\boldsymbol{k},3}-a_{\boldsymbol{k},3}^{\dagger}a_{\boldsymbol{k},1})\boldsymbol{\epsilon}(\boldsymbol{k},2),\\
\boldsymbol{s}_{\boldsymbol{k},3} & =(a_{\boldsymbol{k},2}^{\dagger}a_{\boldsymbol{k},1}-a_{\boldsymbol{k},1}^{\dagger}a_{\boldsymbol{k},2})\boldsymbol{\epsilon}(\boldsymbol{k},3).\label{eq:sk3}
\end{align}
Now, we show that the counter-rotating wave terms in the photon spin also vanish,
\begin{align}
&\frac{i\hbar}{2} \sum_{\lambda\lambda'}\int d^3k a_{\boldsymbol{k},\lambda}a_{-\boldsymbol{k},\lambda'}\boldsymbol{\epsilon}(\boldsymbol{k},\lambda)\times\boldsymbol{\epsilon}(-\boldsymbol{k},\lambda') \nonumber\\
= & \frac{i\hbar}{4} \sum_{\lambda\lambda'}\int d^3k \left[ a_{\boldsymbol{k},\lambda}a_{-\boldsymbol{k},\lambda'} + a_{-\boldsymbol{k},\lambda'}a_{\boldsymbol{k},\lambda}\right]\boldsymbol{\epsilon}(\boldsymbol{k},\lambda)\times\boldsymbol{\epsilon}(-\boldsymbol{k},\lambda')\nonumber\\
= & \frac{i\hbar}{4} \sum_{\lambda\lambda'}\int d^3k  a_{\boldsymbol{k},\lambda}a_{-\boldsymbol{k},\lambda'}[\boldsymbol{\epsilon}(\boldsymbol{k},\lambda)\times\boldsymbol{\epsilon}(-\boldsymbol{k},\lambda')\nonumber\\
& +\boldsymbol{\epsilon}(-\boldsymbol{k},\lambda')\times\boldsymbol{\epsilon}(\boldsymbol{k},\lambda)]=0
\end{align}
Similar argument also applies to the terms $a^{\dagger}_{\boldsymbol{k}\lambda},a^{\dagger}_{-\boldsymbol{k},\lambda'}$.

The Maxwell OAM can be expanded as\begin{widetext}
\begin{align}
\boldsymbol{L}_{M}= & \frac{1}{\mu_{0}c}\int d^{3}x(\partial_{0}A^{\mu})\boldsymbol{x}\times\boldsymbol{\nabla}A_{\mu}=-\frac{1}{c}\int d^{3}x\pi_{\mu}\boldsymbol{x}\times\boldsymbol{\nabla}A^{\mu}\\
= & -\frac{\hbar}{2(2\pi)^{3}}\!\int\!\! d^{3}x\!\int\!\! d^{3}k\int\!\! d^{3}k'\!\sum_{\lambda',\lambda=0}^{3}\epsilon_{\mu}(\boldsymbol{k},\lambda)\epsilon^{\mu}(\boldsymbol{k}',\lambda')\sqrt{\frac{\omega_{\boldsymbol{k}}}{\omega_{\boldsymbol{k}'}}}\left[a_{\boldsymbol{k},\lambda}e^{i\boldsymbol{k}\cdot \boldsymbol{x}}(\boldsymbol{x}\times\boldsymbol{k}')a_{\boldsymbol{k}',\lambda'}^{\dagger}e^{-i\boldsymbol{k}'\cdot\boldsymbol{x} }+a_{\boldsymbol{k},\lambda}^{\dagger}e^{-i\boldsymbol{k}\cdot \boldsymbol{x}}(\boldsymbol{x}\times\boldsymbol{k}')a_{\boldsymbol{k}',\lambda'}e^{i\boldsymbol{k}'\cdot\boldsymbol{x} }\right]\nonumber\\
= & \frac{\hbar}{2(2\pi)^{3}}\!\int\!\! d^{3}x\!\int\!\! d^{3}k\!\int\!\! d^{3}k'\!\sum_{\lambda,\lambda'=0}^{3}\epsilon_{\mu}(\boldsymbol{k},\lambda)\epsilon^{\mu}(\boldsymbol{k}',\lambda')\sqrt{\frac{\omega_{\boldsymbol{k}}}{\omega_{\boldsymbol{k}'}}}\left[a_{\boldsymbol{k},\lambda}a_{\boldsymbol{k}',\lambda'}^{\dagger}e^{i\boldsymbol{k}\cdot \boldsymbol{x}}(\boldsymbol{k}'\times\boldsymbol{x})e^{-i\boldsymbol{k}'\cdot\boldsymbol{x} }+a_{\boldsymbol{k},\lambda}^{\dagger}a_{\boldsymbol{k}',\lambda'}e^{-i\boldsymbol{k}'\cdot \boldsymbol{x}}(\boldsymbol{k}'\times\boldsymbol{x})e^{i\boldsymbol{k}\cdot\boldsymbol{x} }\right]\nonumber\\
= & \frac{i\hbar}{2(2\pi)^{3}}\!\int\!\! d^{3}x\!\int\!\! d^{3}k\!\int\!\! d^{3}k'\!\sum_{\lambda,\lambda'=0}^{3}\epsilon_{\mu}(\boldsymbol{k},\lambda)\epsilon^{\mu}(\boldsymbol{k}',\lambda')\sqrt{\frac{\omega_{\boldsymbol{k}}}{\omega_{\boldsymbol{k}'}}}\left[a_{\boldsymbol{k},\lambda}a_{\boldsymbol{k}',\lambda'}^{\dagger}e^{i\boldsymbol{k}\cdot \boldsymbol{x}}(\boldsymbol{k}'\times\boldsymbol{\nabla}_{\boldsymbol{k}'})e^{-i\boldsymbol{k}'\cdot\boldsymbol{x} }-a_{\boldsymbol{k},\lambda}^{\dagger}a_{\boldsymbol{k}',\lambda'}e^{-i\boldsymbol{k}\cdot \boldsymbol{x}}\boldsymbol{k}'\times\boldsymbol{\nabla}_{\boldsymbol{k}'}e^{i\boldsymbol{k}'\cdot\boldsymbol{x} }\right]\nonumber\\
= & \textcolor{red}{i\hbar\int d^{3}k\sum_{\lambda,\lambda'=0}^{3}
 a_{\boldsymbol{k},\lambda}^{\dagger}\left[
 g_{\lambda\lambda'}(\boldsymbol{k}\times\boldsymbol{\nabla}_{\boldsymbol{k}})
 +\boldsymbol{\Gamma}_{\lambda\lambda'}(\boldsymbol{k})\right]
 a_{\boldsymbol{k},\lambda'}}. \label{eq:PWE-LM}
\end{align}
\end{widetext}
{\color{red}
In the integration by parts leading to Eq.~(\ref{eq:PWE-LM}), the angular momentum derivative acts on the \emph{entire} momentum-dependent factor
$\omega_{\boldsymbol{k}'}^{-1/2}\epsilon^{\mu}(\boldsymbol{k}',\lambda')a_{\boldsymbol{k}',\lambda'}$.
Since $(\boldsymbol{k}\times\boldsymbol{\nabla}_{\boldsymbol{k}})\omega_{\boldsymbol{k}}^{\alpha}=0$, the frequency normalization makes no contribution, whereas the derivative of the polarization frame gives
\begin{equation}
\boldsymbol{\Gamma}_{\lambda\lambda'}(\boldsymbol{k})
\equiv \epsilon_{\mu}(\boldsymbol{k},\lambda)
\bigl[(\boldsymbol{k}\times\boldsymbol{\nabla}_{\boldsymbol{k}})
\epsilon^{\mu}(\boldsymbol{k},\lambda')\bigr].
\label{eq:PWE-LM-connection}
\end{equation}
Equation~(\ref{eq:PWE-LM}) uses the real polarization vectors of the present expansion.
For a momentum-dependent transverse/longitudinal frame, this connection does not generally vanish: for example,
$\boldsymbol{e}_{\theta}\cdot\partial_{\varphi}\boldsymbol{e}_{\varphi}=-\cos\theta$.
The connection-free expression is valid in a fixed internal Cartesian basis, or as shorthand when the momentum derivative is understood to be frame covariant; it is not valid with an ordinary derivative acting on amplitudes in a moving polarization frame.

The counter-rotating terms proportional to $a_{\boldsymbol{k},\lambda}a_{\boldsymbol{k}',\lambda'}$ and to $a_{\boldsymbol{k},\lambda}^{\dagger}a_{\boldsymbol{k}',\lambda'}^{\dagger}$ also cancel \emph{without} a rotating-wave approximation. The spatial integration restricts these terms to $\boldsymbol{k}'=-\boldsymbol{k}$. Symmetrizing under the simultaneous interchange $(\boldsymbol{k},\lambda)\leftrightarrow(\boldsymbol{k}',\lambda')$, and retaining the momentum derivatives of both the amplitudes and the polarization vectors, gives mutually cancelling contributions by bosonic commutativity and the antisymmetry of the angular derivative (with vanishing momentum-space boundary terms).
}

Utilizing the generators of the SO(3) rotation group, we can re-express the photon spin in the same form of the Dirac spin
\begin{equation}
\boldsymbol{S}_{M}=\hbar\int d^{3}k\phi_{\boldsymbol{k}}^{\dagger}\hat{\boldsymbol{s}}\phi_{\boldsymbol{k}},
\end{equation}
where the column vector $\phi_{\boldsymbol{k}}=[a_{\boldsymbol{k},1},a_{\boldsymbol{k},2},a_{\boldsymbol{k},3}]^{T}$
is the field operator of the Maxwell field in wave-vector space and the $3\times 3$ matrix
$\hat{\boldsymbol{s}}=\sum_{\lambda=1}^{3}\hat{s}_{\lambda}\boldsymbol{\epsilon}(\boldsymbol{k},\lambda)$
is the spin operator of the Maxwell field with
\begin{equation}
\hat{s}_{1}=\left[\begin{array}{ccc}
0 & 0 & 0\\
0 & 0 & -i\\
0 & i & 0
\end{array}\right],\ \hat{s}_{2}=\left[\begin{array}{ccc}
0 & 0 & i\\
0 & 0 & 0\\
-i & 0 & 0
\end{array}\right],\ \hat{s}_{3}=\left[\begin{array}{ccc}
0 & -i & 0\\
i & 0 & 0\\
0 & 0 & 0
\end{array}\right]
\end{equation}
satisfying the commutation relation $[\hat{s}_i,\hat{s}_j]=i\varepsilon_{ijk}\hat{s}_k$. 

In Chap. 2 of the textbook~\cite{jauch2012theory}, the authors defined the four-vector photon spin operator as the quantum Stokes parameter operators
\begin{align}
\Sigma_0 & = \int d^3 k (a_{\boldsymbol{k},1}^{\dagger}a_{\boldsymbol{k},1} + a_{\boldsymbol{k},2}^{\dagger}a_{\boldsymbol{k},2}),\\
\Sigma_1 & = \int d^3 k (a_{\boldsymbol{k},1}^{\dagger}a_{\boldsymbol{k},2} + a_{\boldsymbol{k},2}^{\dagger}a_{\boldsymbol{k},1}),\\
\Sigma_2 & = i\int d^3 k (a_{\boldsymbol{k},2}^{\dagger}a_{\boldsymbol{k},1} - a_{\boldsymbol{k},1}^{\dagger}a_{\boldsymbol{k},2}),\\
\Sigma_3 & = \int d^3 k (a_{\boldsymbol{k},1}^{\dagger}a_{\boldsymbol{k},1} - a_{\boldsymbol{k},2}^{\dagger}a_{\boldsymbol{k},2}).
\end{align}
However, this definition has two serious problems. Firstly, none of these four operators is an integer-spin operator, because an extra factor 2 exists in the commutation relations, i.e, $[\Sigma_i,\Sigma_j]=2i\varepsilon_{ijk}\Sigma_k$. Secondly, the direction of this ``photon spin" is completely undetermined, because they are constructed in a phase space instead of the real spacetime. This is significantly different from our defined photon spin operators or the Dirac spin operators as shown in the following. 

In the following, we will use the expansion of the electric field $\boldsymbol{E} =-c\left(\partial^{0}\boldsymbol{A}+\boldsymbol{\nabla}A^{0}\right)=c\mu_{0}\boldsymbol{\pi}-c\boldsymbol{\nabla}A^{0}$ and the magnetic field $\boldsymbol{B}=\boldsymbol{\nabla}\times\boldsymbol{A}$,
\begin{align}
\boldsymbol{E}(\boldsymbol{x}) & =i\int d^{3}k\sqrt{\frac{\hbar\omega_{\boldsymbol{k}}}{2\varepsilon_{0}(2\pi)^{3}}} \left\{[a_{\boldsymbol{k},1}\boldsymbol{\epsilon}(\boldsymbol{k},1)+a_{\boldsymbol{k},2}\boldsymbol{\epsilon}(\boldsymbol{k},2)\right.\nonumber\\
& \left.+(a_{\boldsymbol{k},3}-a_{\boldsymbol{k},0})\boldsymbol{\epsilon}(\boldsymbol{k},3)]e^{i\boldsymbol{k}\cdot \boldsymbol{x}}-{\rm h.c.}\right\}, \label{eq:E_planewave}\\
\boldsymbol{B}(\boldsymbol{x}) & =\frac{i}{c}\int d^{3}k\sqrt{\frac{\hbar\omega_{\boldsymbol{k}}}{2\varepsilon_{0}(2\pi)^{3}}}\nonumber\\
& \times\left\{\left[a_{\boldsymbol{k},1}\boldsymbol{\epsilon}(\boldsymbol{k},2)-a_{\boldsymbol{k},2}\boldsymbol{\epsilon}(\boldsymbol{k},1)\right]e^{i\boldsymbol{k}\cdot \boldsymbol{x}}-{\rm h.c.}\right\}.\label{eq:B_planewave}
\end{align}

\section{Angular commutation relations revisit \label{sec:CR-recheck}}
Utilizing the commutation relations of the ladder
operators $[a_{\boldsymbol{k},\lambda},a_{\boldsymbol{k}',\lambda'}^{\dagger}]=-g_{\lambda\lambda'}\delta^{3}(\boldsymbol{k}-\boldsymbol{k}')$, we now recheck the commutation relations of $\boldsymbol{S}_{M}$
and $\boldsymbol{L}_{M}$. We start from the photon spin
\begin{align}
[S_{M,i},S_{M,j}] \!\!& =\!\!-\hbar^{2}\!\!\int \!\!d^{3}k\!\!\int\!\! d^{3}k'[s_{\boldsymbol{k},\lambda},s_{\boldsymbol{k}',\lambda'}]\epsilon_{i}(\boldsymbol{k},\lambda)\textcolor{red}{\epsilon_{j}(\boldsymbol{k}',\lambda')}\\
& =-\hbar^{2}\!\!\int\!\! d^{3}k\varepsilon_{\lambda\lambda'\lambda^{\prime\prime}}\epsilon_{i}(\boldsymbol{k},\lambda)\epsilon_{j}(\boldsymbol{k},\lambda')s_{\boldsymbol{k},\lambda^{\prime\prime}},
\end{align}
where we have used the expansion $\boldsymbol{e}_{i}=\epsilon_{i}(\boldsymbol{k},\lambda)\boldsymbol{\epsilon}(\boldsymbol{k},\lambda)$ ($\lambda = 1,2,3$) of the unit vectors $\boldsymbol{e}_{i}\ (i=x,y,z)$ of a local fixed coordinate with
the $\boldsymbol{k}$-dependent polarization vectors $\boldsymbol{\epsilon}(\boldsymbol{k},\lambda)$ and the relation $[s_{\boldsymbol{k},\lambda},s_{\boldsymbol{k}',\lambda'}]=\delta^{3}(\boldsymbol{k}-\boldsymbol{k}')\varepsilon_{\lambda\lambda'\lambda^{\prime\prime}}s_{\boldsymbol{k},\lambda^{\prime\prime}}$ for $s_{\boldsymbol{k},\lambda}$ in Eqs.~(\textcolor{red}{\ref{eq:sk1}--\ref{eq:sk3}}).
The cross product of two unit vectors will give
\begin{align}
\left(\boldsymbol{e}_{i}\times\boldsymbol{e}_{j}\right)_{\lambda^{\prime\prime}} = \varepsilon_{\lambda\lambda'\lambda^{\prime\prime}}\epsilon_{i}(\boldsymbol{k},\lambda)\epsilon_{j}(\boldsymbol{k},\lambda')=\varepsilon_{ijl}\epsilon_l(\boldsymbol{k},\lambda'').
\end{align}
Then, we obtain the stand commutation relation for SAM of the Maxwell field
\begin{equation}
[S_{M,i},S_{M,j}]=-\hbar^{2}\int d^{3}k\varepsilon_{ijl}s_{\boldsymbol{k},\lambda^{\prime\prime}}\epsilon_{l}(\boldsymbol{k},\lambda^{\prime\prime})=i\hbar\varepsilon_{ijl}S_{M,l}.
\end{equation}

Similarly, we check the OAM commutation relation\textcolor{red}{, retaining the polarization-frame connection of Eq.~(\ref{eq:PWE-LM-connection})}.
{\color{red}
For the complete orthonormal spatial polarization triad, introduce the derivative
\begin{equation}
(\mathcal D_a)_{rs}=\delta_{rs}\partial_{k_a}
+\boldsymbol\epsilon(\boldsymbol k,r)\cdot
 \partial_{k_a}\boldsymbol\epsilon(\boldsymbol k,s),\qquad r,s=1,2,3.
\end{equation}
Equation~(\ref{eq:PWE-LM}) then takes the form
\begin{align}
L_{M,i}= & i\hbar\int d^3k\,a_{\boldsymbol k,0}^\dagger
(\boldsymbol k\times\boldsymbol\nabla_{\boldsymbol k})_i a_{\boldsymbol k,0} \nonumber \\
& -i\hbar\int d^3k\sum_{r,s=1}^{3}a_{\boldsymbol k,r}^\dagger
(\boldsymbol k\times\boldsymbol{\mathcal D})_{i,rs}a_{\boldsymbol k,s}.
\end{align}
The complete triad implies $[\mathcal D_a,\mathcal D_b]=0$ and
$[\mathcal D_a,k_b]=\delta_{ab}$; in particular,
\begin{equation}
[(\boldsymbol k\times\boldsymbol{\mathcal D})_i,
 (\boldsymbol k\times\boldsymbol{\mathcal D})_j]
=-\varepsilon_{ijk}(\boldsymbol k\times\boldsymbol{\mathcal D})_k,
\end{equation}
with the same identity for $\boldsymbol{\mathcal D}$ replaced by $\boldsymbol\nabla_{\boldsymbol k}$ in the scalar sector. Using
$[a_{\boldsymbol k,0},a_{\boldsymbol k',0}^{\dagger}]=-\delta^3(\boldsymbol k-\boldsymbol k')$
and $[a_{\boldsymbol k,r},a_{\boldsymbol k',s}^{\dagger}]=\delta_{rs}\delta^3(\boldsymbol k-\boldsymbol k')$, we obtain
\begin{equation}
[L_{M,i},L_{M,j}]=i\hbar\varepsilon_{ijk}L_{M,k}.\label{eq:LM1}
\end{equation}
}

Finally, we show that $\boldsymbol{S}_M$ and $\boldsymbol{L}_M$ commute with each other.
{\color{red}
Recall from Appendix~\ref{sec:plane-wave expansion} that
$S_{M,i}=\hbar\int d^3k\,\phi_{\boldsymbol k}^{\dagger}
\hat s_i(\boldsymbol k)\phi_{\boldsymbol k}$, where
$\hat s_i(\boldsymbol k)=\sum_{\lambda=1}^{3}
\epsilon_i(\boldsymbol k,\lambda)\hat s_{\lambda}$.
For the complete spatial triad, this spin matrix is covariantly constant:
\begin{equation}
[\mathcal D_a,\hat s_i(\boldsymbol k)]=0.
\end{equation}
The scalar part of $L_{M,j}$ commutes with $S_{M,i}$ separately;
thus the full canonical operators satisfy
\begin{equation}
[S_{M,i},L_{M,j}]=0.
\end{equation}
The momentum derivative of the polarization vectors is essential to this cancellation.
}

\section{Gauge-invariant angular momentum observables\label{sec:appendixD}}
In this section, we show how to recover the gauge invariance of the QED angular momentum. To fully solve this problem, we start from the Lagrangian density of the combined Dirac-Maxwell fields,
\begin{equation}
\mathcal{L}_{\rm QED}\!=\!i\hbar c\bar{\psi}\gamma^{\mu}\partial_{\mu}\psi\!-\!mc^{2}\bar{\psi}\psi\!-\!\frac{1}{2\mu_{0}}(\partial_{\mu}A^{\nu})(\partial^{\mu}A_{\nu}) \!-\!qc\bar{\psi}\gamma_{\mu}A^{\mu}\psi,
\end{equation}
where $\psi$ is the Dirac field operator, $\bar{\psi}=\psi^{\dagger}\gamma^{0}$, and
\begin{equation}
\gamma^{0}=\beta,\ \gamma^{i}=\beta\alpha_{i},\ i=1,2,3
\end{equation}
with
\begin{equation}
\beta=\left[\begin{array}{cc}
I & 0\\
0 & -I
\end{array}\right],\ \alpha_{i}=\left[\begin{array}{cc}
0 & \sigma_{i}\\
\sigma_{i} & 0
\end{array}\right],
\end{equation}
the $2\times2$ identity matrix $I$, and the Pauli matrices $\sigma_{i}$.

From the Noether's theorem, we obtain the total angular momentum of the system $\boldsymbol{J}=\boldsymbol{S}_{D}+\boldsymbol{L}_{D}+\boldsymbol{S}_{M}+\boldsymbol{L}_{M}$. The SAM and OAM of the Dirac field have been well studied and understood
\begin{align}
\boldsymbol{S}_{D} & =\frac{1}{2}\hbar\int d^{3}x\psi^{\dagger}\hat{\boldsymbol{\Sigma}}\psi,\\
\boldsymbol{L}_{D} & =\int d^{3}x\psi^{\dagger}\boldsymbol{x}\times\boldsymbol{p}\psi, \label{eq:S_D}
\end{align}
where 
\begin{equation}
\hat{\boldsymbol{\Sigma}}=\left[\begin{array}{cc}
\hat{\boldsymbol{\sigma}} & 0\\
0 & \hat{\boldsymbol{\sigma}}
\end{array}\right],\label{eq:L_D}
\end{equation}
is the Dirac spin operator. Utilizing the anti-commutation rules of the Dirac field,
\begin{align}
[\psi_{r}(\boldsymbol{x},t),\psi_{r'}^{\dagger}(\boldsymbol{x}',t)]_{+} & =\delta_{rr'}\delta^{3}(\boldsymbol{x}-\boldsymbol{x}'), \label{eq:ETCR-D1} \\{}
[\psi_{r}(\boldsymbol{x},t),\psi_{r'}(\boldsymbol{x}',t)]_{+} & =[\psi_{r}^{\dagger}(\boldsymbol{x},t),\psi_{r'}^{\dagger}(\boldsymbol{x}',t)]_{+}=0,
\end{align}
we can verify the commutation relations for the angular momentum of the Dirac field\begin{align}
[S_{D,i},S_{D,j}] & =i\hbar\varepsilon_{ijk}S_{D,k},\\{}
[L_{D,i},L_{D,j}] & =i\hbar\varepsilon_{ijk}L_{D,k},\\{}
[L_{D,i},S_{D,j}] & =0. \label{eq:CR-LD}
\end{align}
We can easily check that the angular momenta of the Dirac and Maxwell fields commute with each other, because the quantum operators of these two fields act on different Hilbert spaces.

To obtain the gauge-invariant parts of the SAM and OAM of photons, we split both the vector potential $\boldsymbol{A}=\boldsymbol{A}_{\perp}+\boldsymbol{A}_{\parallel}$ and the canonical momentum operator $\boldsymbol{\pi}=\boldsymbol{\pi}_{\perp}+\boldsymbol{\pi}_{\parallel}$ into transverse and longitudinal parts~\cite{chen2008spin}, where 
\begin{align}
\boldsymbol{\nabla}\cdot\boldsymbol{A}_{\perp} & =0,\ \boldsymbol{\nabla}\times\boldsymbol{A}_{\parallel}=0.\\
\boldsymbol{\nabla}\cdot\boldsymbol{\pi}_{\perp} & =0,\ \boldsymbol{\nabla}\times\boldsymbol{\pi}_{\parallel}=0.
\end{align}

Then, the total photon spin can be split into three parts
\begin{equation}
\boldsymbol{S}_{M}=\frac{1}{c}\int d^{3}x[\boldsymbol{\pi}_{\perp}\times\boldsymbol{A}_{\perp}+\boldsymbol{\pi}_{\parallel}\times\boldsymbol{A}_{\perp}+\boldsymbol{\pi}_{\perp}\times\boldsymbol{A}_{\parallel}],    
\end{equation}
where the contribution from $\boldsymbol{\pi}_{\parallel}\times\boldsymbol{A}_{\parallel}$ is zero. The gauge-invariant part of the photon spin is
\begin{equation}
\boldsymbol{S}^{\rm obs}_{M}\!=\!\frac{1}{c}\!\!\int\!\! d^{3}x\boldsymbol{\pi}_{\perp}\times\boldsymbol{A}_{\perp}\!=\!i\hbar\!\!\int\!\! d^{3}k\left(a_{\boldsymbol{k},2}^{\dagger}a_{\boldsymbol{k},1}\!-\!a_{\boldsymbol{k},1}^{\dagger}a_{\boldsymbol{k},2}\right)\boldsymbol{\epsilon}(\boldsymbol{k},3).
\end{equation}

Similarly, the total OAM of the photon can be split into
\begin{equation}
\begin{aligned}
\boldsymbol{L}_{M}& = \frac{1}{c}\int d^{3}x\,[\pi_{\perp}^{j}\boldsymbol{x}\times\boldsymbol{\nabla}A_{\perp}^{j}
+\pi_{\parallel}^{j}\boldsymbol{x}\times\boldsymbol{\nabla}A_{\parallel}^{j}\\
&-\pi_{0}\boldsymbol{x}\times\boldsymbol{\nabla}A_{0}
\textcolor{red}{+\pi_{\perp}^{j}\boldsymbol{x}\times\boldsymbol{\nabla}A_{\parallel}^{j}+\pi_{\parallel}^{j}\boldsymbol{x}\times\boldsymbol{\nabla}A_{\perp}^{j}}].
\end{aligned}
\end{equation}
and its gauge-invariant
part is given by\textcolor{red}{
\begin{equation}
\begin{aligned}
&\boldsymbol{L}^{\rm obs}_{M} \!=\!\frac{1}{c}\int d^{3}x\,\pi_{\perp}^{j}\boldsymbol{x}\times\boldsymbol{\nabla}A_{\perp}^{j}\\
&\!=\!-i\hbar\int d^{3}k\sum_{\lambda=1,2}
 a_{\boldsymbol{k},\lambda}^{\dagger}(\boldsymbol{k}\times\boldsymbol{D}_{\boldsymbol{k},\lambda})a_{\boldsymbol{k}},
\end{aligned}
\end{equation}
where $\bm D_{\boldsymbol{k},\lambda}=\bm\nabla_{\bm k}-i\bm{\mathcal A}_{\boldsymbol{k},\lambda}$ with Berry connection
\begin{equation}
 \bm{\mathcal A}_{\boldsymbol{k},\lambda} = i\,\bm\epsilon^*(\boldsymbol k,\lambda)\!\cdot\!\bm\nabla_{\bm k}\bm\epsilon (\boldsymbol k,\lambda).\label{eq:PWE-Lperp-connection}  
\end{equation}}
Using the relations between the transverse part of $\boldsymbol{\pi}$
and $\boldsymbol{E}$, we can rewrite $\boldsymbol{S}^{\rm obs}_{M}$
and $\boldsymbol{L}^{\rm obs}_{M}$ as
\begin{align}
\boldsymbol{S}^{\rm obs}_{M} &=\varepsilon_{0}\int d^{3}x\boldsymbol{E}_{\perp}\times\boldsymbol{A}_{\perp}, \label{eq:SMp}\\
\boldsymbol{L}^{\rm obs}_{M} & =\varepsilon_{0}\int d^{3}xE_{\perp}^{j}\boldsymbol{x}\times\boldsymbol{\nabla}A_{\perp}^{j},
\end{align}
which reduce to the angular momentum of the classical transverse EM field exactly.

The gauge-invariant OAM of the Dirac field is given by
\begin{equation}
 \boldsymbol{L}^{\rm obs}_{D} =\boldsymbol{L}_{D}+\boldsymbol{L}_{M}+\boldsymbol{S}_{M} -\boldsymbol{L}^{\rm obs}_{M}-\boldsymbol{S}^{\rm obs}_{M}\equiv \boldsymbol{L}_{D} +\boldsymbol{L}_{\rm pure}, \label{eq:LDp-L}
\end{equation}
where 
\begin{equation}
\begin{aligned}
\boldsymbol{L}_{\rm pure}=&\frac{1}{c}\int d^{3}x\,[\boldsymbol{\pi}_{\parallel}\times\boldsymbol{A}_{\perp}
+\boldsymbol{\pi}_{\perp}\times\boldsymbol{A}_{\parallel}+\textcolor{red}{\pi_{\perp}^{j}\boldsymbol{x}\times\boldsymbol{\nabla}A_{\parallel}^{j}}\\
& +\textcolor{red}{\pi_{\parallel}^{j}\boldsymbol{x}\times\boldsymbol{\nabla}A_{\perp}^{j}}+\pi_{\parallel}^{j}\boldsymbol{x}\times\boldsymbol{\nabla}A_{\parallel}^{j}-\pi_{0}\boldsymbol{x}\times\boldsymbol{\nabla}A_{0}].
\end{aligned} 
\end{equation} 
is a pure-gauge contribution. We note that $\boldsymbol{L}_{\rm pure}$ contains the OAM of both longitudinal and scalar photons.

{\color{red}
The mixed spin and orbital terms cancel pairwise up to surface integrals. To see this directly, write
$\boldsymbol{A}_{\parallel}=\boldsymbol{\nabla}\chi$ and
$\boldsymbol{\pi}_{\parallel}=\boldsymbol{\nabla}\varphi$ and use
$\boldsymbol{\nabla}\cdot\boldsymbol{\pi}_{\perp}=
\boldsymbol{\nabla}\cdot\boldsymbol{A}_{\perp}=0$.
Integration by parts yields
\begin{align}
\int d^3x\,\pi_{\perp}^{j}\boldsymbol{x}\times\boldsymbol{\nabla}A_{\parallel}^{j}
&=\int d^3x\,\pi_{\perp}^{j}\boldsymbol{x}\times\partial_j\boldsymbol{\nabla}\chi =-\int d^3x\,\boldsymbol{\pi}_{\perp}\times\boldsymbol{A}_{\parallel},
\label{eq:L_PS1}\\
\int d^3x\,\pi_{\parallel}^{j}\boldsymbol{x}\times\boldsymbol{\nabla}A_{\perp}^{j}
&=\int d^3x\,(\partial_j\varphi)\boldsymbol{x}\times\boldsymbol{\nabla}A_{\perp}^{j}\nonumber\\
=\int d^3x & \,\varphi(\boldsymbol{\nabla}\times\boldsymbol{A}_{\perp})=-\int d^3x\,\boldsymbol{\pi}_{\parallel}\times\boldsymbol{A}_{\perp}.
\label{eq:L_PS2}
\end{align}
This cancellation involves the \emph{sum} of the mixed spin and orbital contributions.
}

\textcolor{red}{Thus, only the diagonal longitudinal and scalar contributions remain. Their polarization-frame derivatives give no additional diagonal term, since $\boldsymbol{\epsilon}(\boldsymbol{k},3)\cdot[(\boldsymbol{k}\times\boldsymbol{\nabla}_{\boldsymbol{k}})\boldsymbol{\epsilon}(\boldsymbol{k},3)]=0$ and the scalar polarization is momentum independent.}
\begin{align}
\boldsymbol{L}_{\rm pure} = i\hbar\!\!\int\!\! d^{3}k\left(a_{\boldsymbol{k},0}^{\dagger}\boldsymbol{k}\times\boldsymbol{\nabla}_{\boldsymbol{k}}a_{\boldsymbol{k},0}\!-\!a_{\boldsymbol{k},3}^{\dagger}\boldsymbol{k}\times\boldsymbol{\nabla}_{\boldsymbol{k}}a_{\boldsymbol{k},3}\right). \label{eq:L_pure}
\end{align}
Next, we show how to remove the gauge-dependence in $\boldsymbol{L}^{\rm obs}_D$ by enforcing the Gupta-Bleuler gauge constraint.

\subsection{Gupta-Bleuler condition in the Lorenz gauge}
To guarantee that the Lagrangian $\mathcal{L}_{\rm QED}$ gives the correct motion equations, we need to add the gauge constraint on the four-potential $A_{\mu}$. In classical electrodynamics, the Lorenz condition $\partial^{\mu}A_{\mu}=0$ has been applied~\cite{jackson1999classical}. However, this gauge condition can not be generalized as an operator identity directly. We can easily verify that~\cite{jauch2012theory}
\begin{equation}
[\partial^{\mu}A_{\mu}(\boldsymbol{x},t),A_{\nu}(\boldsymbol{x}',t)]=i\hbar c\mu_{0}g_{0,\nu}\delta^{3}(\boldsymbol{x}-\boldsymbol{x}')\neq0.
\end{equation}
Thus, $\partial^{\mu}A_{\mu}$ can not be a zero operator. This problem has been solved by Gupta and Bleuler independently~\cite{gupta1950theory,bleuler1950new}, by enforcing the following constraint for all physical state $|\Phi\rangle$
\begin{equation}
 \partial^{\mu}A_{\mu}^{(+)}|\Phi\rangle = 0,\ \langle\Phi|\partial^{\mu}A_{\mu}^{(-)} = 0,  
\end{equation}
where $A_{\mu}^{(+)}$ and $A_{\mu}^{(-)}$ are the positive and negative frequency parts of $A_{\mu}$, respectively. The summation of the positive and negative frequency parts recovers the classical Lorenz-gauge condition,
\begin{equation}
\langle\Phi|\partial_{\mu} A^{\mu}|\Phi\rangle = \langle\Phi|( \partial^{\mu}A_{\mu}^{(+)}+ \partial^{\mu}A_{\mu}^{(-)})|\Phi\rangle  =0 \label{eq:GB-full_lorenz}.  
\end{equation} 
Thus, the Gupta-Bleuler condition is the quantum version of the Lorenz gauge condition. Bleuler has also generalized the upper constraint to the case when the EM field is coupled to a charge. However, as shown in the Chap. V of~\cite{cohen1997photons}, a more straightforward way is to calculate the Heisenberg equation for $A_0^{(+)}$ after performing the plane-wave expansion of $A_{\mu}$. 

The full Hamiltonian describing the interaction of Dirac-Maxwell fields in the Lorenz gauge is given by~\cite{cohen1997photons}
\begin{equation}
H=H_{D}+H_{M}^{T}+H_{M}^{L}+H_{M}^{S}+H_{{\rm int}}^{T}+H_{{\rm int}}^{L}+H_{{\rm int}}^{S},
\end{equation}
with the Dirac Hamiltonian
\begin{equation}
H_{D}=\int d^{3}x\psi^{\dagger}(\boldsymbol{x},t)\left( c\boldsymbol{\alpha}\cdot\boldsymbol{p}+\beta mc^{2}\right)\psi(\boldsymbol{x},t),
\end{equation}
the transverse, longitudinal, and scalar modes of the Maxwell field
\begin{align}
H_{M}^{T} & =\int\hbar\omega_{\boldsymbol{k}}\left(a_{\boldsymbol{k},1}^{\dagger}a_{\boldsymbol{k},1}+a_{\boldsymbol{k},2}^{\dagger}a_{\boldsymbol{k},2}\right)d^{3}k,\\
H_{M}^{L} & =\int\hbar\omega_{\boldsymbol{k}}a_{\boldsymbol{k},3}^{\dagger}a_{\boldsymbol{k},3}d^{3}k,\\
H_{M}^{S} & =-\int\hbar\omega_{\boldsymbol{k}}a_{\boldsymbol{k},0}^{\dagger}a_{\boldsymbol{k},0}d^{3}k.
\end{align}
Using the definitions of the charge density and current operators,
\begin{align}
\rho_{e}(\boldsymbol{x}) & =q\psi^{\dagger}(\boldsymbol{x})\psi(\boldsymbol{x}),\\
\boldsymbol{j}_{e}(\boldsymbol{x}) & =qc\psi^{\dagger}(\boldsymbol{x})\boldsymbol{\alpha}\psi(\boldsymbol{x}),
\end{align}
the interaction parts are given by
\begin{align}
H_{{\rm int}}^{T}+H_{{\rm int}}^{L} & =-\int d^{3}x\boldsymbol{j}_e(\boldsymbol{x})\cdot\boldsymbol{A}(\boldsymbol{x})\nonumber\\
\!\!\!\!\!&=\!-\!\int\!\! d^3 k\hbar\omega_{\boldsymbol{k}}\sum_{\lambda=1}^{3}\left[a^{\dagger}_{\boldsymbol{k},\lambda}\boldsymbol{\xi}(\boldsymbol{k})\cdot\boldsymbol{\epsilon}(\boldsymbol{k},\lambda) +\rm{h.c.}\right],\\
H_{{\rm int}}^{S} & = c\int d^3x \rho_e(\boldsymbol{x})A_0(\boldsymbol{x})\nonumber\\
& =\int d^{3}k\hbar\omega_{\boldsymbol{k}}\left[\xi_0(\boldsymbol{k})a_{\boldsymbol{k},0}^{\dagger}+\xi_0^{*}(\boldsymbol{k})a_{\boldsymbol{k},0}\right],\label{eq:Hint_S}
\end{align}
where 
\begin{align}
\xi_0(\boldsymbol{k}) & =\frac{c}{\hbar\omega_{\boldsymbol{k}}}\sqrt{\frac{\hbar}{2\varepsilon_{0}\omega_{\boldsymbol{k}}(2\pi)^{3}}}\int d^{3}x\rho_{e}(\boldsymbol{x})e^{-i\boldsymbol{k}\cdot\boldsymbol{x}} \label{eq:xi} \\
\boldsymbol{\xi}(\boldsymbol{k}) & =\frac{1}{\hbar\omega_{\boldsymbol{k}}}\sqrt{\frac{\hbar}{2\varepsilon_{0}\omega_{\boldsymbol{k}}(2\pi)^{3}}}\int d^{3}x\boldsymbol{j}_e(\boldsymbol{x})e^{-i\boldsymbol{k}\cdot\boldsymbol{x}}.
\end{align}

We now give the Gupta-Bleuler condition for the coupled Dirac-Maxwell fields. In the Heisenberg picture, the motion equation of the scalar field is given by
\begin{equation}
\dot{a}_{\boldsymbol{k},0}=\frac{i}{\hbar}[H,a_{\boldsymbol{k},0}]=-i\omega_{\boldsymbol{k}}[a_{\boldsymbol{k},0}-\xi_0(\boldsymbol{k})].
\end{equation}
Here, we see that the time-dependence of the scalar annihilation operator does not follow the free-field one $a_{\boldsymbol{k},0}(t)\neq a_{\boldsymbol{k},0}(0)\exp^{-i\omega_{\boldsymbol{k}}t}$, due to the coupling to the Dirac field. 

The Gupta-Bleuler condition requires that the four-divergence of the positive frequency part of $A_{\mu}$ acting on any physical state $|\Phi\rangle$ equals zero. To hold for all plane-wave modes, this requires~\cite{cohen1997photons}
\begin{equation}
\left[\frac{1}{c}\dot{a}_{\boldsymbol{k},\lambda}+i|\boldsymbol{k}|a_{\boldsymbol{k},3}\right]\left|\Phi\right\rangle =i|\boldsymbol{k}|\left[a_{\boldsymbol{k},3}-a_{\boldsymbol{k},0}+\xi_0(\boldsymbol{k})\right]\left|\Phi\right\rangle =0,
\end{equation}
i.e.,
\begin{equation}
\left[a_{\boldsymbol{k},3}-a_{\boldsymbol{k},0}+\xi_0(\boldsymbol{k})\right]\left|\Phi\right\rangle =0. \label{eq:Gupta1}
\end{equation}
We emphasize that the Gupta-Bleuler condition for the combined system in the Lorenz gauge is different from the free-space one~\cite{gupta1950theory,greiner2013field}, which do not have the shift $\xi_0(\boldsymbol{k})$.

Applying the Gupta-Bleuler constraint (\ref{eq:Gupta1}) and its Hermitian conjugate to Eq.~(\ref{eq:L_pure}), we have\begin{widetext}
\begin{align}
\langle\Phi|\boldsymbol{L}_{\rm pure}|\Phi\rangle & = i\hbar\int d^{3}k\left\{\langle\Phi|[a_{\boldsymbol{k},3}^{\dagger}+\xi_0^{*}(\boldsymbol{k})]\left(\boldsymbol{k}\times\boldsymbol{\nabla}_{\boldsymbol{k}}\right)[a_{\boldsymbol{k},3}+\xi_0(\boldsymbol{k})]|\Phi\rangle-\langle\Phi|\xi_0(\boldsymbol{k})\left(\boldsymbol{k}\times\boldsymbol{\nabla}_{\boldsymbol{k}}\right)a_{\boldsymbol{k},3}^{\dagger}|\Phi\rangle\right\} \\
& = i\hbar\int d^{3}k\left\{\langle\Phi|\left[\xi_0^{*}(\boldsymbol{k})\left(\boldsymbol{k}\times\boldsymbol{\nabla}_{\boldsymbol{k}}\right)a_{\boldsymbol{k},3}-\xi_0(\boldsymbol{k})\left(\boldsymbol{k}\times\boldsymbol{\nabla}_{\boldsymbol{k}}\right)a_{\boldsymbol{k},3}^{\dagger}\right]|\Phi\rangle+\langle\Phi|\xi_0(\boldsymbol{k})(\boldsymbol{k}\times\boldsymbol{\nabla}_{\boldsymbol{k}})\xi_0^{*}(\boldsymbol{k})|\Phi\rangle\right\}.  
\end{align}\end{widetext}
The last term vanishes due to the identity,
\begin{equation}
\xi_0(-\boldsymbol{k})(-\boldsymbol{k}\times\boldsymbol{\nabla}_{-\boldsymbol{k}})\xi_0^{*}(-\boldsymbol{k}) = \xi_0^*(\boldsymbol{k})(\boldsymbol{k}\times\boldsymbol{\nabla}_{\boldsymbol{k}})\xi_0(\boldsymbol{k})  
\end{equation}
where we have used the fact $\xi_0(-\boldsymbol{k})= \xi_0^*(\boldsymbol{k})$.
Using the following plane-wave expansion,\begin{widetext}
\begin{align} 
-q\int d^{3}x\psi^{\dagger}\boldsymbol{x}\times\boldsymbol{A}_{\parallel}\psi
= & -\int d^{3}x\int d^{3}k\sqrt{\frac{\hbar}{2\varepsilon_{0}\omega_{\boldsymbol{k}}(2\pi)^{3}}}\rho_{e}(\boldsymbol{x})\boldsymbol{x}\times\left[a_{\boldsymbol{k},3}\boldsymbol{\epsilon}(\boldsymbol{k},3)e^{i\boldsymbol{k}\cdot\boldsymbol{x} }+a_{\boldsymbol{k},3}^{\dagger}\boldsymbol{\epsilon}(\boldsymbol{k},3)e^{-i\boldsymbol{k}\cdot\boldsymbol{x} }\right]\\
= & \int d^{3}x\int d^{3}k\sqrt{\frac{\hbar}{2\varepsilon_{0}\omega_{\boldsymbol{k}}(2\pi)^{3}}}\rho_{e}(\boldsymbol{x})\left[a_{\boldsymbol{k},3}\boldsymbol{\epsilon}(\boldsymbol{k},3)\times\boldsymbol{x}e^{i\boldsymbol{k}\cdot\boldsymbol{x} }+a_{\boldsymbol{k},3}^{\dagger}\boldsymbol{\epsilon}(\boldsymbol{k},3)\times\boldsymbol{x}e^{-i\boldsymbol{k}\cdot\boldsymbol{x} }\right]\\
= & \int d^{3}x\int d^{3}k\sqrt{\frac{\hbar}{2\varepsilon_{0}\omega_{\boldsymbol{k}}(2\pi)^{3}}}\rho_{e}(\boldsymbol{x})\left[a_{\boldsymbol{k},3}\frac{\boldsymbol{k}}{|\boldsymbol{k}|}\times(-i\boldsymbol{\nabla}_{\boldsymbol{k}})e^{i\boldsymbol{k}\cdot\boldsymbol{x} }+a_{\boldsymbol{k},3}^{\dagger}\frac{\boldsymbol{k}}{|\boldsymbol{k}|}\times(i\boldsymbol{\nabla}_{\boldsymbol{k}})e^{-i\boldsymbol{k}\cdot\boldsymbol{x} }\right]\\
= & i\int d^{3}x\int d^{3}k\sqrt{\frac{\hbar}{2\varepsilon_{0}\omega_{\boldsymbol{k}}(2\pi)^{3}}}\rho_{e}(\boldsymbol{x})\left[e^{i\boldsymbol{k}\cdot\boldsymbol{x} }\frac{\boldsymbol{k}}{|\boldsymbol{k}|}\times\boldsymbol{\nabla}_{\boldsymbol{k}}a_{\boldsymbol{k},3}-e^{-i\boldsymbol{k}\cdot\boldsymbol{x} }\frac{\boldsymbol{k}}{|\boldsymbol{k}|}\times\boldsymbol{\nabla}_{\boldsymbol{k}}a_{\boldsymbol{k},3}^{\dagger}\right]\\
= & i\hbar\int d^{3}k\left[\xi_0^{*}(\boldsymbol{k})\left(\boldsymbol{k}\times\boldsymbol{\nabla}_{\boldsymbol{k}}\right)a_{\boldsymbol{k},3}-\xi_0(\boldsymbol{k})\left(\boldsymbol{k}\times\boldsymbol{\nabla}_{\boldsymbol{k}}\right)a_{\boldsymbol{k},3}^{\dagger}\right],
\end{align}\end{widetext}
we have 
\begin{equation}
\langle\Phi|\boldsymbol{L}_{\rm pure}|\Phi\rangle = \langle\Phi|-q\int d^{3}x\psi^{\dagger}\boldsymbol{x}\times\boldsymbol{A}_{\parallel}\psi|\Phi\rangle,    
\end{equation}
which guarantees that the mean value $\langle\Phi|\boldsymbol{L}^{\rm obs}_D|\Phi\rangle$ is gauge invariant.

\section{Commutation relations for the observables\label{sec:appendixE}}

In Table~\ref{tab:gauge}, we summarize the commutation relations between the angular momenta of the QED system for both canonical and gauge-invariant decompositions. In this section, we focus on the commutation relations for the gauge-invariant decomposition of the total QED angular momentum, which is given by: $\boldsymbol{J}=\boldsymbol{L}^{\rm obs}_{D}+\boldsymbol{S}_{D}+\boldsymbol{L}^{\rm obs}_{M}+\boldsymbol{S}^{\rm obs}_{M}$. 

It is easily to check that
\begin{equation}
[S^{\rm obs}_{M,i},S^{\rm obs}_{M,j}]=0.
\end{equation}
Because the photon spin for plane-wave modes $\boldsymbol{s}_{\boldsymbol{k},3} =(a_{\boldsymbol{k},2}^{\dagger}a_{\boldsymbol{k},1}-a_{\boldsymbol{k},1}^{\dagger}a_{\boldsymbol{k},2})\boldsymbol{\epsilon}(\boldsymbol{k},3)$ with different $\boldsymbol{k}$ commutes, i.e., $[\boldsymbol{s}_{\boldsymbol{k},3},\boldsymbol{s}_{\boldsymbol{k}',3}]=0$. For a single plane wave, the three components of $\boldsymbol{s}_{\boldsymbol{k},3}$ in a local coordinate frame also commute with other.

{\color{red}
Unlike the full canonical OAM, the observable transverse OAM contains the polarization-frame connection in Eq.~(\ref{eq:PWE-Lperp-connection}). Its helicity curvature $\boldsymbol{\nabla}_{\boldsymbol{k}}\times\boldsymbol{\mathcal A}_{\boldsymbol{k},h}=-h\hat{\boldsymbol{k}}/|\boldsymbol{k}|^{2}$ ($h=\pm $) gives
\begin{equation}
[L_{M,i}^{\rm obs},L_{M,j}^{\rm obs}]
=i\hbar\varepsilon_{ijk}\bigl(L_{M,k}^{\rm obs}-S_{M,k}^{\rm obs}\bigr).
\label{eq:LM-lambda}
\end{equation}
Furthermore, $[(\boldsymbol{k}\times\boldsymbol{D}_{\boldsymbol{k},h})_{j},\hat{k}_{i}]=(\boldsymbol{k}\times\boldsymbol{\nabla}_{\boldsymbol{k}})_{j}\hat{k}_{i}=\varepsilon_{ijk}\hat{k}_{k}$, so the spin--OAM cross commutator is
\begin{equation}
[S^{\rm obs}_{M,i},L^{\rm obs}_{M,j}]
=i\hbar\varepsilon_{ijk}S^{\rm obs}_{M,k}.
\end{equation}
Together with $[S^{\rm obs}_{M,i},S^{\rm obs}_{M,j}]=0$, these relations ensure that $\boldsymbol{J}^{\rm obs}_{M}=\boldsymbol{L}^{\rm obs}_{M}+\boldsymbol{S}^{\rm obs}_{M}$ obeys the standard SO(3) angular-momentum algebra.
}
Since $\boldsymbol{L}^{\rm obs}_{D}$ does not contain transverse Maxwell
modes, then we can easily obtain
\begin{equation}
[L_{D,i}^{\rm obs},S^{\rm obs}_{M,j}]=[L_{D,i}^{\rm obs},L_{M,j}^{\rm obs}]=0.
\end{equation}

\textcolor{red}{From Eqs.~(\ref{eq:CR-LD}) and (\ref{eq:L_pure}), the scalar/longitudinal contribution satisfies $[L_{{\rm pure},i},L_{{\rm pure},j}]=i\hbar\varepsilon_{ijk}L_{{\rm pure},k}$ and commutes with $\boldsymbol{L}_{D}$. Hence, using Eq.~(\ref{eq:LDp-L}),}
\begin{equation}
[L_{D,i}^{\rm obs},L_{D,j}^{\rm obs}]=i\hbar\varepsilon_{ijk}L_{D,k}^{\rm obs}.
\end{equation}

\section{Angular momentum operators from the standard QED Lagrangian in Coulomb gauge \label{sec:appendixF}}
The modern gauge field theory for QED is based on the gauge invariance of the standard Lagrangian density~\cite{Pauli1941relativistic,yang1954conservation,griffiths2008introduction}
\begin{equation}
\mathcal{L}_{\rm QED,ST}\!=\!i\hbar c\bar{\psi}\gamma^{\mu}\partial_{\mu}\psi\!-\!mc^{2}\bar{\psi}\psi \!-\!qc\bar{\psi}\gamma_{\mu}A^{\mu}\psi \!-\!\frac{1}{4\mu_0}F^{\mu\nu}F_{\mu\nu}. \label{eq:LaQED_ST}  
\end{equation}
In this subsection, we show how to obtain the gauge-invariant decomposition of the angular momentum from the standard QED Lagrangian.

Following the Neother's theorem, Jaffe and Manohar have given a decomposition of the total angular momentum of QED $\boldsymbol{J}=\boldsymbol{S}_D+\boldsymbol{L}_{D}+\boldsymbol{S}_{M,{\rm JM}}+\boldsymbol{L}_{M,{\rm JM}}$~\cite{jaffe1990g1,leader2014angular}.  The SAM and OAM of the Maxwell field are given by
\begin{equation}
\boldsymbol{S}_{M,{\rm JM}}  = \varepsilon_{0}\int d^{3}x \boldsymbol{E}\times\boldsymbol{A},
\end{equation}
\begin{equation}
\boldsymbol{L}_{M,{\rm JM}}  = \varepsilon_{0}\int d^{3}xE^{j}\boldsymbol{x}\times\boldsymbol{\nabla}A^{j}.
\end{equation}

Similar to the canonical decomposition obtained from the Lorenz gauge, $\boldsymbol{L}_D$, $\boldsymbol{S}_{M,{\rm JM}}$, and $\boldsymbol{L}_{M,{\rm JM}}$ are not gauge invariant. There is another problem that the longitudinal part of the electric field can not be quantized. As explain in Chap. II of the text book~\cite{cohen1997photons}, both the scalar potential $A_0$ and the longitudinal vector potential $\boldsymbol{A}_{\parallel}$ are redundant dynamical variables, which can be eliminated via the Euler-Lagrange equation for $A_0$ and the Coulomb gauge condition $\boldsymbol{\nabla}\cdot\boldsymbol{A}=0$ (i.e., $\boldsymbol{A}_{\parallel}=0$). The reduced QED Lagrangian in the Coulomb gauge is given by
\begin{align}
L'_{\rm QED,ST} \!=& \!\!\int\!\! d^3 x \left\{\textcolor{red}{i\hbar c\bar{\psi}\gamma^{\mu}\partial_{\mu}\psi}\!-\!mc^{2}\bar{\psi}\psi \!-\!\int\!\! d^{3}x'\frac{\rho_{e}(\boldsymbol{x})\rho_{e}(\boldsymbol{x}')}{8\pi\varepsilon_{0}|\boldsymbol{x}-\boldsymbol{x}'|}\right.\nonumber \\ &\!\!\!\!\!\!\! \!\!\!\!\!\!\! \left.-qc\psi^{\dagger}\boldsymbol{\alpha}\cdot\boldsymbol{A}_{\perp}\psi\!+\!\frac{1}{2\mu_0}[(\partial_0\boldsymbol{A}_{\perp})^2\!-\!(\boldsymbol{\nabla}\times\boldsymbol{A}_{\perp})^2]\right\}\!.
\end{align}
The quantization of the Dirac-Maxwell fields is actually based on this reduced Lagrangian density.

The quantization procedure of the Dirac field does not change. The canonical momentum of the EM field is given by~\cite{greiner2013field,cohen1997photons}
\begin{equation}
\boldsymbol{\pi}_{\perp} = \frac{1}{\mu_0}\partial_0\boldsymbol{A}_{\perp}= - \frac{1}{c\mu_0}\boldsymbol{E}_{\perp}.
\end{equation}
The quantization of the EM field in the Coulomb gauge can be achieved by postulating the following commutation relation,
\begin{align}
[A_{\perp}^i(\boldsymbol{x},t),E_{\perp}^j(\boldsymbol{x}',t)]=\textcolor{red}{-i\frac{\hbar }{\varepsilon_0}\delta_{\perp}^{ij}(\boldsymbol{x}-\boldsymbol{x}')},    
\end{align}
where 
\begin{equation}
\delta_{\perp}^{ij}(\boldsymbol{x}-\boldsymbol{x}') = \frac{1}{(2\pi)^3}\int d^3k e^{i\boldsymbol{k}\cdot\boldsymbol{x}}\left(\delta^{ij}-\frac{k_i k_j}{|\boldsymbol{k}|^2}\right)
\end{equation}
is the transverse delta function.

In the Coulomb gauge, the longitudinal part of the quantum field operator $\boldsymbol{A}$ vanishes, i.e., $\boldsymbol{A}_{\parallel}=0$. The OAM angular momentum automatically reduces to our defined gauge-invariant OAM of the photon,
\begin{equation}
\boldsymbol{L}_{M,{\rm JM}}  = \boldsymbol{L}'_M = \varepsilon_{0}\int d^{3}xE^{j}_{\perp}\boldsymbol{x}\times\boldsymbol{\nabla}A^{j}_{\perp}.
\end{equation}
By splitting both the electric field $\boldsymbol{E}$ and the vector potential $A$ into transverse and longitudinal parts, we have
\begin{equation}
\boldsymbol{S}_{M,\rm{JM}} =\varepsilon_0 \int d^3 x\left[\boldsymbol{E}_{\perp}\times\boldsymbol{A}_{\perp} + \boldsymbol{E}_{\perp}\times\boldsymbol{A}_{\parallel}+\boldsymbol{E}_{\parallel}\times\boldsymbol{A}_{\perp} \right],
\end{equation}
where the second term vanishes in the Coulomb gauge. Using the relations
\begin{align}
\boldsymbol{E}_{\parallel}\times\boldsymbol{A}_{\perp} & =(\boldsymbol{E}_{\parallel}\cdot\boldsymbol{\nabla})\boldsymbol{x}\times\boldsymbol{A}_{\perp}-\boldsymbol{x}\times(\boldsymbol{E}_{\parallel}\cdot\boldsymbol{\nabla})\boldsymbol{A}_{\perp},
\end{align}
and integral by parts, we have
\begin{align}
\varepsilon_0\int d^3 x &\boldsymbol{E}_{\parallel}\times\boldsymbol{A}_{\perp} \nonumber\\
= & - \varepsilon_0\int d^3 x \left[ (\boldsymbol{\nabla}\cdot\boldsymbol{E}_{\parallel})\boldsymbol{x}\times\boldsymbol{A}_{\perp}+\boldsymbol{x}\times(\boldsymbol{E}_{\parallel}\cdot\boldsymbol{\nabla})\boldsymbol{A}_{\perp}\right], \label{eq:SM_parallel1}
\end{align}
where we have neglected a boundary term during the partial integral. Now, we use plane-wave expansion to verify that the two terms in (\ref{eq:SM_parallel1}) actually cancel out. Since the longitudinal electric field $\boldsymbol{E}_{\parallel}(\boldsymbol{x})$ has not been quantized in Coulomb gauge, we expand $\boldsymbol{E}_{\parallel}(\boldsymbol{x})$ 
\begin{equation}
\boldsymbol{E}_{\parallel}(\boldsymbol{x})=i\int d^{3}k\sqrt{\frac{\hbar\omega_{\boldsymbol{k}}}{2\varepsilon_{0}(2\pi)^{3}}}\left(\alpha_{\boldsymbol{k}}e^{i\boldsymbol{k}\cdot\boldsymbol{x}}-\alpha_{\boldsymbol{k}}^{*}e^{-i\boldsymbol{k}\cdot\boldsymbol{x}}\right)\frac{\boldsymbol{k}}{|\boldsymbol{k}|},
\end{equation}
with classical functions $\alpha_{-\boldsymbol{k}}=\alpha_{\boldsymbol{k}}^{*}$. Using the same techniques in evaluating Eqs.~(\ref{eq:L_PS1}) and (\ref{eq:L_PS2}), we have
\begin{align}
& \varepsilon_{0}\int d^{3}x(\boldsymbol{\nabla}\cdot\boldsymbol{E}_{\parallel})\boldsymbol{x}\times\boldsymbol{A}_{\perp} = -\varepsilon_{0}\int d^{3}x\boldsymbol{x}\times(\boldsymbol{E}_{\parallel}\cdot\boldsymbol{\nabla})\boldsymbol{A}_{\perp}\nonumber \\=&i\hbar\sum_{\lambda=1,2}\!\int\!\! d^{3}k|\boldsymbol{k}|\left[\alpha_{\boldsymbol{k}}\boldsymbol{\nabla}_{\boldsymbol{k}}\!\times\!\boldsymbol{\epsilon}(\boldsymbol{k},\lambda)a_{\boldsymbol{k},\lambda}^{\dagger}\!-\!\alpha_{\boldsymbol{k}}^{*}\boldsymbol{\nabla}_{\boldsymbol{k}}\!\times\!\boldsymbol{\epsilon}(\boldsymbol{k},\lambda)a_{\boldsymbol{k},\lambda}\right].
\end{align}

Finally, we obtain the gauge-invariant decomposition of the QED angular momentum $\boldsymbol{J}=\boldsymbol{L}+\boldsymbol{S}_{D}+\boldsymbol{L}^{\rm obs}_{M}+\boldsymbol{S}^{\rm obs}_{M}$, which recovers the results obtained in the Lorenz quantization frame-work. We note that in the Coulomb gauge, the pure gauge contribution to the OAM of the Dirac field disappears.

\section{Contrast with previous decompositions\label{sec:appendixG}}
In the review article~\cite{leader2014angular}, the authors have listed another five decompositions of the QED angular momentum~\cite{belinfante1939spin,jaffe1990g1,Ji1997gauge,chen2008spin,Wakamatsu2010gauge}, \textcolor{red}{which represent the same total angular momentum after the appropriate boundary terms and Gauss-law constraints are accounted for, although the separate contributions differ.} Some of the decompositions did not separate the SAM and OAM of the photon~\cite{belinfante1939spin,Ji1997gauge}. \textcolor{red}{The remaining decompositions distinguish canonical and kinetic matter OAM and distribute the Gauss-law-dependent terms between matter and electromagnetic contributions~\cite{jaffe1990g1,chen2008spin,Wakamatsu2010gauge}. The commutation properties of the individual contributions must therefore be checked in the appropriate constrained Hilbert space.} In Table~\ref{tab:3}, we contrast our decomposition of the QED angular momentum with previous results.

In the following, we show some problems about the commutation relations in previous decompositions. \textcolor{red}{The longitudinal electric field is fixed by the operator Gauss law, $\boldsymbol\nabla\cdot\boldsymbol E=\rho_e/\varepsilon_0$, rather than being an independent radiative degree of freedom (see Appendix~\ref{sec:appendixF}). The Lorenz-gauge free-field expansions in Eqs.~(\ref{eq:E_planewave}) and (\ref{eq:B_planewave}) can be used to inspect the free-field sector, but cannot by themselves establish the commutators of interacting matter--field contributions.}

\subsection{The Belinfante and Ji decompositions \label{sec:BJ-decomposition}}

In the Belinfante and Ji decompositions, the electromagnetic angular momentum is not separated into spin and orbital contributions. Its commutation algebra must therefore be examined within the full Lorenz-gauge quantization framework, retaining the transverse, longitudinal, and scalar modes before imposing the physical-state condition.

{\color{red}
The Belinfante electromagnetic angular momentum is defined by the gauge-invariant operator
\begin{equation}
\boldsymbol J_{M}
=\varepsilon_0\int d^3x\,
 \boldsymbol x\times(\boldsymbol E\times\boldsymbol B).
\label{eq:JM-BJ}
\end{equation}
Here, both $\boldsymbol E$ and $\boldsymbol B$ are constructed from the same four-component Lorenz-gauge potential. Using the canonical equal-time commutators given above, without imposing any subsidiary condition, we obtain
\begin{align}
[E_a(\boldsymbol x),E_b(\boldsymbol y)]
&=[B_a(\boldsymbol x),B_b(\boldsymbol y)]=0,\nonumber\\
[E_a(\boldsymbol x),B_b(\boldsymbol y)]
&=-\frac{i\hbar}{\varepsilon_0}\varepsilon_{abc}
 \partial_{x^c}\delta^3(\boldsymbol x-\boldsymbol y).
\end{align}
Substituting these relations into Eq.~\eqref{eq:JM-BJ} and integrating by parts yields
\begin{align}
[J_{M,i},J_{M,j}]
=i\hbar\varepsilon_{ijk}\Bigg[J_{M,k}
+\varepsilon_0\int d^3x\,x_k
 (\boldsymbol x\cdot\boldsymbol B)(\nabla\cdot\boldsymbol E)\Bigg].
\label{eq:JM-BJ-CR}
\end{align}
In the presence of charges, Gauss's law,
$\varepsilon_0\nabla\cdot\boldsymbol E=\rho_e$,
implies that the additional term on the right-hand side of Eq.~\eqref{eq:JM-BJ-CR} does not generally vanish.
}

\subsection{The Jaffe--Manohar decomposition \label{sec:JM-decomposition}}
The Jaffe--Manohar decomposition reads
$\boldsymbol J=\boldsymbol S_D+\boldsymbol L_D+
\boldsymbol S_{M,{\rm JM}}+\boldsymbol L_{M,{\rm JM}}$~\cite{jaffe1990g1,leader2014angular},
where the SAM and OAM of the Maxwell field are given by
\begin{equation}
\boldsymbol S_{M,{\rm JM}}=\varepsilon_0\int d^3x\,
\boldsymbol E\times\boldsymbol A,
\end{equation}
\begin{equation}
\boldsymbol L_{M,{\rm JM}}=\varepsilon_0\int d^3x\,
E^j\boldsymbol x\times\boldsymbol\nabla A^j,
\end{equation}
respectively. This decomposition is known to be gauge non-invariant~\cite{leader2014angular}.
\textcolor{red}{Nevertheless, gauge dependence should not be confused with a failure of the formal commutation algebra. We evaluate the latter in the full Lorenz-gauge field space, without first discarding scalar or longitudinal contributions.}

{\color{red}
The electric field contains the scalar--longitudinal combination, whereas the spatial potential $\boldsymbol A$ contains both transverse and longitudinal modes. Rather than inferring commutators from a truncated momentum-space expansion, we work directly with the complete equal-time field algebra. The unprojected Lorenz-gauge equal-time field commutators give
\begin{align}
[A_a(\boldsymbol x),E_b(\boldsymbol y)]
&=-\frac{i\hbar}{\varepsilon_0}\delta_{ab}
\delta^3(\boldsymbol x-\boldsymbol y),\nonumber\\
[A_a(\boldsymbol x),A_b(\boldsymbol y)]
&=[E_a(\boldsymbol x),E_b(\boldsymbol y)]=0.
\label{eq:JM-field-ETCR}
\end{align}
Here $a,b=1,2,3$ refer to fixed Cartesian components. These relations have not been projected onto the transverse subspace. For $X_j=A_j,E_j$,
\begin{align}
[S_{M,{\rm JM},i},X_j(\boldsymbol x)]
 &=i\hbar\varepsilon_{ijk}X_k(\boldsymbol x),\nonumber\\
[L_{M,{\rm JM},i},X_j(\boldsymbol x)]
 &=i\hbar (\boldsymbol x\times\boldsymbol\nabla)_iX_j(\boldsymbol x).
\end{align}
Thus the spin and orbital parts generate, respectively, rotations of the Cartesian vector indices and rotations of the spatial arguments. Using the same equal-time commutators, with normal ordering and vanishing boundary terms understood, we obtain
\begin{align}
[S_{M,{\rm JM},i},S_{M,{\rm JM},j}]
 &=i\hbar\varepsilon_{ijk}S_{M,{\rm JM},k},\nonumber\\
[L_{M,{\rm JM},i},L_{M,{\rm JM},j}]
 &=i\hbar\varepsilon_{ijk}L_{M,{\rm JM},k},\nonumber\\
[S_{M,{\rm JM},i},L_{M,{\rm JM},j}]&=0.
\label{eq:JM-CR-cov}
\end{align}
These are formal identities for the \emph{unprojected} equal-time operators; they do not assert that the separate gauge-dependent operators are observables on the physical-state quotient, nor do they replace the full four-component covariant Noether generator. In particular, the null-mode commutator
$[a_{\boldsymbol k,3}-a_{\boldsymbol k,0},
 a^\dagger_{\boldsymbol k',3}-a^\dagger_{\boldsymbol k',0}]=0$
does not permit the scalar and longitudinal modes to be deleted before evaluating Eq.~\eqref{eq:JM-field-ETCR}. An explicit momentum-space evaluation in a $\boldsymbol k$-dependent polarization frame must also retain the polarization-frame connection (Appendix~\ref{sec:plane-wave expansion}).
}

\subsection{The Chen et al. and the Wakamatsu decompositions \label{sec:chen}}
To solve the gauge dependent problem, Chen \textit{et al}. split the gauge field  $\boldsymbol{A}$ into physical (transverse) and pure-gauge (longitudinal) parts, i.e., $\boldsymbol{A}=\boldsymbol{A}_{\perp}+\boldsymbol{A}_{\parallel}$. Then, they put the gauge dependent parts in $\boldsymbol{S}_{M,{\rm JM}}$ and $\boldsymbol{L}_{M,{\rm JM}}$ into $\boldsymbol{L}_{D}$. Finally, they obtained the ``guage-invariant" decomposition of the the QED angular momentum $\boldsymbol{J}=\boldsymbol{S}_D+\boldsymbol{L}_{D,{\rm Chen}}+\boldsymbol{S}_{M,{\rm Chen}}+\boldsymbol{L}_{M,{\rm Chen}}$, where the OAM of the Dirac field, SAM, and OAM of the Maxwell field are given by
\begin{equation}
\boldsymbol{L}_{D,{\rm Chen}}= \int d^{3}x\left[-i\hbar\psi^{\dagger}\boldsymbol{x}\times\boldsymbol{\nabla}\psi-\textcolor{red}{\rho_e}\boldsymbol{x}\times\boldsymbol{A}_{\parallel}\right], \label{eq:LDchen1}
\end{equation}
\begin{equation}
\boldsymbol{S}_{M,{\rm Chen}} =\varepsilon_{0}\int d^{3}x\boldsymbol{E}\times\boldsymbol{A}_{\perp},
\end{equation}
\begin{equation}
\boldsymbol{L}_{M,{\rm Chen}} =\varepsilon_{0}\int d^{3}xE^{j}\boldsymbol{x}\times\boldsymbol{\nabla}A_{\perp}^{j}.
\end{equation}
\textcolor{red}{The matter OAM and the free-field spin expansion, together with the exact orbital decomposition, are}
\begin{align}
\boldsymbol{L}_{D,{\rm Chen}} &\textcolor{red}{=\int d^3x\,\psi^\dagger\boldsymbol{x}\times
 \bigl(-i\hbar\boldsymbol\nabla-q\boldsymbol A_\parallel\bigr)\psi}.
\label{eq:LDchen}
\end{align}
\begin{align}
\boldsymbol{S}_{M,{\rm Chen}} &=\frac{i\hbar}{2}\int d^{3}k\left[(a_{\boldsymbol{k},3}^{\dagger}-a_{\boldsymbol{k},0}^{\dagger})a_{\boldsymbol{k},2}\boldsymbol{\epsilon}(\boldsymbol{k},1)\right.\nonumber\\
&\left.\!\!\!\!\!\!\!+a_{\boldsymbol{k},1}^{\dagger}(a_{\boldsymbol{k},3}\!-\!a_{\boldsymbol{k},0})\boldsymbol{\epsilon}(\boldsymbol{k},2)\!+\!2a_{\boldsymbol{k},2}^{\dagger}a_{\boldsymbol{k},1}\boldsymbol{\epsilon}(\boldsymbol{k},3)\!-\!{\rm h.c.}\right]\!, \label{eq:SMchen}\!
\end{align}
\begin{align}
\boldsymbol{L}_{M,{\rm Chen}}
&\textcolor{red}{=\boldsymbol L_M^{\rm obs}
+\varepsilon_0\int d^3x\,E_\parallel^j
 \boldsymbol x\times\boldsymbol\nabla A_\perp^j}\\
&\textcolor{red}{=\boldsymbol L_M^{\rm obs}
-\varepsilon_0\int d^3x\,\boldsymbol E_\parallel\times\boldsymbol A_\perp},
\end{align}

\textcolor{red}{The Chen matter OAM contains only the pure-gauge potential $\boldsymbol A_\parallel=\boldsymbol\nabla\chi$. Since $\boldsymbol\nabla\times\boldsymbol A_\parallel=0$, its orbital components obey the standard algebra (assuming the usual boundary/domain conditions), rather than the inequality obtained from an unconstrained null-mode calculation:}
{\color{red}
\begin{equation}
[L^i_{D,{\rm Chen}},L^j_{D,{\rm Chen}}]
=i\hbar\varepsilon^{ijk}L^k_{D,{\rm Chen}}.
\end{equation}
}
\textcolor{red}{The photon SAM and OAM above contain opposite longitudinal-electric-field terms; these cancel in their sum (Appendix~\ref{sec:appendixF}). In the source-free transverse sector their separate commutators are instead those of Appendix~\ref{sec:appendixE}; in particular $[\boldsymbol S_M^{\rm obs},\boldsymbol L_M^{\rm obs}]\ne0$. For interacting fields, commutators involving $\boldsymbol E_\parallel$ must be evaluated with the operator Gauss law, not by treating the longitudinal field as an independent free photon.}

Similar issues also exist in Wakamatsu decomposition. \textcolor{red}{With the charge-density term retained, the Wakamatsu decomposition is} $\boldsymbol{J}=\boldsymbol{S}_D+\boldsymbol{L}_{D,{\rm Wak}}+\boldsymbol{S}_{M,{\rm Wak}}+\boldsymbol{L}_{M,{\rm Wak}}$, where
\begin{equation}
\boldsymbol{L}_{D,{\rm Wak}} = \int d^{3}x\left[-i\hbar\psi^{\dagger}\boldsymbol{x}\times\boldsymbol{\nabla}\psi-\textcolor{red}{\rho_e}\boldsymbol{x}\times\boldsymbol{A}\right], \label{eq:LDWak}
\end{equation}
\begin{equation}
\boldsymbol{S}_{M,{\rm Wak}} =\varepsilon_{0}\int d^{3}x\boldsymbol{E}\times\boldsymbol{A}_{\perp},
\end{equation}
\begin{equation}
\boldsymbol{L}_{M,{\rm Wak}} =\varepsilon_{0}\int d^{3}x\,\textcolor{red}{\left[E^{j}\boldsymbol{x}\times\boldsymbol{\nabla}A_{\perp}^{j}+(\boldsymbol{\nabla}\cdot\boldsymbol{E})\boldsymbol{x}\times\boldsymbol{A}_{\perp}\right]}.
\end{equation}
\textcolor{red}{The kinetic matter OAM and the electromagnetic terms share the charge--field contribution through the Gauss constraint; their individual commutators must be evaluated using the constrained QED operator algebra. The free-field null-mode identity alone does not determine these interacting commutators.} 

\bibliography{main}
\end{document}